\documentclass{aa}  
\usepackage{graphicx}
\usepackage{txfonts}
\usepackage{natbib}
\usepackage{xcolor}
\usepackage{hyperref}
\hypersetup{
    colorlinks=true,
    linktoc=all,  
    citecolor=blue,   
    linkcolor=red,   
    urlcolor=blue      
}
\usepackage[version=4]{mhchem}
\usepackage{booktabs}
\usepackage{array}
\usepackage{longtable}
\usepackage{float}
\usepackage{subfigure}
\usepackage{amsmath}
\usepackage{amssymb}
\usepackage{algorithm}
\usepackage{algorithmic}
\usepackage{tabularx}
\usepackage[utf8]{inputenc}
\usepackage[T1]{fontenc}
\usepackage{textcomp}
\usepackage{gensymb}
\usepackage{siunitx}
\usepackage[export]{adjustbox} 
\usepackage{lastpage}
\usepackage{comment}
\usepackage[nameinlink, capitalise]{cleveref}
\long\def\ce#1{\ensuremath{#1}}

\begin{document} 

   \title{Characterizing the hierarchical structure of filaments}
   \subtitle{   
     I. The origin of the length-mass ($L-M$) scaling relation
   }

   \author{Rahul Patel\inst{1,2}
          \and
          Daniel Seifried\inst{2}
          \and
          Alvaro Hacar\inst{3}
          }

    \institute{
    Hamburger Sternwarte, Universit\"at Hamburg, Gojenbergsweg 112, 21029 Hamburg, Germany\\
    \email{rahul.patel@uni-hamburg.de}
    \and
    I. Physikalisches Institut, Universit\"at zu K\"oln, Z\"ulpicher Strasse 77, 50937, K\"oln, Germany
    \and
    Institute for Astrophysics, University of Vienna, T\"urkenschanzstrasse 17, 1180 Vienna, Austria
    }
   \date{Received xx.xx.xxxx; accepted xx.xx.xxxx}

\abstract
{}
   {Filamentary structures in molecular clouds are widely recognized as fundamental components
   of the star formation process.
   Observations report a scaling relation between the length and mass of these filamentary
   structures of the form $L \propto M^{\alpha}$ with $\alpha \simeq 0.5$.
   We aim to characterize this scaling relation in simulated molecular clouds and to
   investigate the influence of hierarchical fragmentation and spatial resolution on it.
   }
   {We applied a two-step filament identification method in high-resolution, 3D hydrodynamical and magnetohydrodynamical simulations of molecular clouds within the SILCC-Zoom project, combining the Rolling Hough Transform for filament identification with dendrogram-based segmentation for hierarchical decomposition. The resulting structures were analyzed across multiple spatial resolutions to quantify their scaling properties.}
   {The identified filamentary structures cover a wide range of column densities ($\sim10^{21}$ to $\sim10^{23}$ cm$^{-2}$). The ensemble of structures, spanning both multiple resolutions and, separately, the full hierarchical decomposition at the highest resolution, exhibits a sub-linear $L-M$ relation close to $L \propto M^{0.5}$. This indicates that this relation is an intrinsic property of hierarchically structured clouds rather than an artifact of resolution blending. When individual structures are traced across different resolutions, the power-law indices remain sub-linear ($\alpha \simeq 0.45 - 0.70$), matching the geometric prediction $L \propto M^{0.5}$ of a fragmentation-driven random walk. However, by decomposing individual structures at the highest resolution, we find a wide range of $L-M$ slopes ($\alpha \simeq 0.2$ to $3$) peaking at $\alpha < 1$. We present a simple toy model that shows that this diversity arises from differences in the internal column density distribution and in the geometry of the segmentation.
   }
   {Taken together, the various results, which all show an average sub-linear power-law relation 
   $L \propto M^{\alpha}$ with $\alpha \simeq 0.5$, indicate that this $L-M$ relation is a robust property of the hierarchical structure and fragmentation occurring in molecular clouds.}


   \maketitle

\section{Introduction}
Stars are born in the densest and coldest phases of the interstellar medium (ISM), specifically within molecular clouds (MCs), where temperatures reach 10 to 15~K and number densities range from $10^2$ to $10^6~\mathrm{cm^{-3}}$ \citep{Chevance_2023}. These clouds are hierarchically structured and host a variety of nested substructures, of which filamentary structures are the most prominent \citep[see the review by][]{hacar_2023}.

Filaments in the ISM were first recognized as elongated dark structures in dust extinction maps \citep{schneider_elmegreen}, and their presence has since been confirmed through \ce{HI} observations \citep{McClure-Griffiths_2006} and \ce{CO} line surveys \citep{bally, schuller_2017, cubuk_2023}, among many others. Although there is no universally accepted definition of a filament, they are generally understood as elongated structures with high aspect ratios,  typically $\gtrsim5$ to $10$,  and significantly enhanced densities relative to their surroundings \citep{schisano_2014, andre_2014}. Their ubiquity and importance were firmly established through far-infrared observations with the Herschel Space Observatory, which revealed intricate networks of filamentary structures in nearly all nearby clouds \citep{andre_2010, andre_2014, molinari_et_al, arzoumanian_2011, koenyves_2015}, present not only in active star-forming regions but also in more diffuse, quiescent environments such as the Polaris translucent cloud \citep{Ward-Thompson_Whitworth_2011}.

The filamentary ISM spans an enormous dynamic range: from sub-parsec, velocity-coherent ``fibers'' \citep{hacar_2018, socci_2024} to structures hundreds of parsecs long \citep{goodman_2014, zucker_2015, zucker_2018, wang_2015, Li_2016, schisano_2014, schisano_2020}, from translucent to optically thick surface densities \citep{arzoumanian_2011}, and from isolated linear strands to complex, twisted formations nested hierarchically within one another \citep[for a comprehensive overview see][]{hacar_2023}; magnetic fields appear to be dynamically important throughout \citep[e.g.,][]{planck_2016, seifried_2020}. This diversity raises the question of whether filaments across scales obey common, possibly universal, scaling relations. Scaling relations have long been central to characterizing the ISM, from the classical relations of \citet{larson_1981} and \citet{heyer_2009}, which connect the size, velocity dispersion, and mass of MCs, to the mass-size relation that \mbox{\citet{kauffmann_2010}} extended from clouds down to dense cores. For filaments specifically, \mbox{\citet{hacar_2023}} established a tight correlation between length and mass,
\begin{equation}
    L \propto M^{\alpha}, \; \mathrm{with} \; \alpha \approx 0.5 \pm 0.2,
\end{equation}
which holds across a wide variety of environments. This sub-linear relation suggests that filaments are approximately self-similar, hierarchically organized structures with a continuously varying line mass, and the physical origin of this sub-linear is the central focus of this work. Its interpretation is not straightforward, however: the measured scaling depends on how structures are identified and how their size is defined, and different size definitions can yield different relations for the same underlying gas distribution \citep{colman_2024}. Projection can likewise alter the slope relative to the intrinsic three-dimensional one. A measured $\alpha \approx 0.5$ therefore does not correspond to a unique physical scenario, which motivates controlled numerical experiments in which hierarchy, resolution, and projection can be varied independently.

Whether this scaling is intrinsic to the filamentary ISM is not settled. Hierarchically organized filament networks emerge self-consistently in modern simulations of the star-forming ISM \citep{zhao_2024, pillworth_2025, pillsworth_2026, koletic_2026}. A sub-linear relation could, however, also arise from the way filaments are observed and identified: compilations combine surveys with different distances and angular resolutions, such that a filament appearing as one long, massive structure at coarse resolution may be resolved into shorter, less massive sub-filaments at higher resolution. Filaments also migrate through the $L-M$ plane as they evolve. Recently, \citet{Feng_2024} obtained $L \propto M^{0.45}$ for simulated filaments pooled across three resolutions and all snapshots (their Fig.~5), and attributed the motion of individual filaments in the plane to accretion, segmentation, and dispersal (their Fig.~10). Our study is complementary: at a fixed evolutionary stage, we isolate the role of hierarchical, nested substructures, together with the role of spatial resolution. This requires us to identify filaments reliably across the full range of scales and column densities they span, including faint substructures.

A wide range of filament-finding algorithms has been developed for this purpose, each with characteristic strengths and limitations:  crest-tracing methods such as \texttt{DisPerSE} \citep{Sousbie_2011}, which provide one-dimensional skeletons but can fragment in faint or high-dynamic-range regions \citep{green_2017, Jiang_2025}; curvature-based Hessian methods, which are intensity-selective and tend to miss faint structures \citep{schisano_2014, salji_2015, planck_int_xxxii_2016}; multi-scale filtering with \texttt{GETFILAMENTS} \citep{menshchikov_2013}; adaptive-threshold skeletonization with \texttt{FilFinder} \citep{koch_2015}; orientation-based detection with   \texttt{FilDReaMS} \citep{carriere_2022}; template-matching approaches \citep[e.g.,][]{juvela_2016}; the \texttt{HiFIVe} algorithm, which uniquely exploits the kinematic information in molecular-line spectral cubes to identify velocity-coherent ``fibers'' in 3D \citep{hacar_2018, socci_2024}; and, more recently, machine-learning techniques \citep{riccio_2016, zavagno_2023, Berthelot_2024}. A common difficulty is that methods operating directly on intensity or column-density maps tend to respond preferentially to the brightest peaks, so faint, low-contrast substructures are merged into larger objects or lost -- precisely the structures whose recovery is essential for probing the hierarchy. 

Motivated by this, we developed a tailored two-step identification strategy in which two techniques play distinct, sequential roles. We first applied the Rolling Hough Transform \citep[RHT;][]{Clark_2014}, a feature-extraction algorithm that is sensitive to coherent linear morphology independent of brightness and therefore captures faint and diffuse filaments on an equal footing with bright ones. We then applied a dendrogram analysis \citep{rosolowsky_2008, astrodendro_2019} to the RHT output in order to decompose the resulting filamentary network into its hierarchy of branches and leaves. In short, the RHT detects the filaments, and the dendrogram resolves their hierarchy, a combination specifically suited to studying how nested substructure shapes the $L-M$ relation.

We applied this method to 3D magnetohydrodynamical (MHD) simulations of MC formation embedded in a  multiphase ISM, taken  from the SILCC-Zoom project \citep{seifried_2017}, which is part of the galactic-scale SILCC project \citep{SILCC, Girichidis_2016}. Using simulations rather than observations allows us to sidestep limitations such as finite spatial resolution, single lines of sight, and projection, and to analyze the same clouds across multiple resolutions and viewing directions. For each identified structure, we measured its length $L$, mass $M$, and mean column density, and studied their distribution in the $L-M$ plane, with particular attention paid to the roles of hierarchical fragmentation and spatial resolution.

The paper is organized as follows. In Sect.~\ref{sec:simulations} we describe the SILCC-Zoom simulations used in this study, including the physical processes incorporated, the adopted resolution, and the subset of simulation snapshots selected for analysis. Section~\ref{sec:methodology} details our two-step identification method -- the RHT, used to detect coherent linear features, and the dendrogram analysis, applied to the RHT output to decompose them into branches and leaves -- together with the measurement of physical properties. In Sect.~\ref{sec:results} we present the $L-M$ relation for  the entire ensemble of filaments, analyzing the relation's slope, its variation across resolutions, and the impact of hierarchical fragmentation upon it. We repeat this analysis at the level of individual filaments in Sect.~\ref{sec:results_individual} and present a toy model to interpret our results. Section~\ref{sec:discussion} discusses the physical implications of our findings and compares them with previous observational and theoretical studies, before we conclude in Sect.~\ref{sec:conclusion}.

\section{Simulations}\label{sec:simulations}
\subsection{Numerics and initial conditions}\label{subsec:numerics}

This study is based on simulations from the SILCC project \citep{SILCC, Girichidis_2016}, extended by the SILCC-Zoom simulations of individual MCs \citep{seifried_2017, Seifried_2019}.  These simulations model the formation and early evolution of MCs within a stratified, supernova-driven multiphase ISM, including self-gravity, magnetic fields (in the MHD runs), a non-equilibrium chemical network for H$_2$ and CO formation, and the associated heating and cooling, on a galactic-disk patch.  The simulations are performed using the adaptive mesh refinement code \texttt{FLASH} \citep{Fryxell_2000, DUBEY2009512}. In the hydrodynamic (HD) case, the MHD Bouchut 5-wave solver \citep{Bouchut_2007, Waagan_2009} is employed with the magnetic field strength set to zero, while the magnetohydrodynamic (MHD) simulations utilize an entropy-stable solver that ensures positive pressure and reduces numerical dissipation \citep{Derigs_2016, Derigs_2017, Derigs_2018}.

The simulations use a domain of size $500~\mathrm{pc}\times500~\mathrm{pc}\times\pm5~\mathrm{kpc}$ with periodic boundary conditions in the $x-$ and $y-$ directions and outflow conditions in the $z-$ direction. The gas is initially distributed according to a vertical Gaussian density profile, $\rho(z)=\rho_0 \exp\left(-z^2/2h_z^2\right)$, with scale height $h_z=30~\mathrm{pc}$ and mid-plane density $\rho_0 = 9\times10^{-24}~\mathrm{g~cm^{-3}}$, which corresponds to a gas surface density of $\Sigma_{\mathrm{gas}} = 10~\mathrm{M_\odot~pc^{-2}}$ \citep{SILCC}. The initial temperature of the gas is $T  = 4500$~K, and for the MHD case, magnetic fields are initialized along the $x$-axis with $B_x(z) = B_{x,0}\sqrt{\rho(z)/\rho_0}$, where $B_{x,0}=3~\mathrm{\mu G}$.

Self-gravity is included via a tree-based solution to Poisson’s equation $\nabla^2\phi=4\pi G\rho$, implemented in \texttt{FLASH} using the method of \citet{wunsch_2018}. An external gravitational potential due to the stellar disk is modeled as a $sech^2$ profile \citep{spitzer_1942} with $\Sigma_\star=30~\mathrm{M_\odot~ pc^{-2}}$ and a scale height of 100~pc.

The simulations include a detailed non-equilibrium chemical network \citep{nelson_1997, glover_2010, glover_clark_2012} that evolves the abundances of \ce{H^+}, \ce{H}, \ce{H_2}, \ce{C^+}, \ce{CO}, \ce{O}, and electrons. Radiative cooling from atomic and molecular species, along with heating from cosmic rays, and via the photoelectric effect, are included. Attenuation of interstellar radiation field (ISRF) is handled using the \texttt{TreeRay} algorithm \citep{clark_glover_klessen_2012, wunsch_2018}, which computes the shielding from dust and H$_2$. The strength of the ISRF is set to 1.7 Habing fields \citep{Habing68,Draine78}.

Turbulence is driven initially by discrete supernova (SN) explosions at a rate consistent with the Kennicutt--Schmidt law. Each SN injects $10^{51}~\mathrm{ergs}$ of energy, using either thermal or momentum injection depending on local resolution \citep{gatto_2015}. Half of the SNe are randomly placed (Gaussian height of 50~pc), and half are located at local density peaks. The SILCC simulations are run at a maximum spatial resolution of 3.9~pc. For further details on the simulation setup, refer to \citet{SILCC}.

\subsection{SILCC-Zoom simulations}\label{subsec:zoom-in}
In the SILCC-Zoom simulations, selected regions  with sizes of $\sim$100~pc, where MCs form, are followed with higher resolution through an adaptive refinement scheme, while the surrounding environment is simultaneously maintained at lower resolution \citep{seifried_2017, Seifried_2019}. The
surrounding ISM remains at the base grid resolution (3.9~pc), while refinement within the zoom region reaches the finest resolution of 0.122~pc. The refinement strategy uses both a second-derivative density criterion \citep{Lohner_1987} and a Jeans refinement criterion \citep{Truelove_1997}, requiring at least 16 cells per Jeans length. The refinement process is started at time $t_0$, where the injection of SNe to drive turbulence is switched off. This ensures that the clouds evolve without significant SNe interactions, ensuring that their subsequent dynamics are governed primarily by gravity, magnetic and thermal pressure, as well as the turbulence inherited from the parent simulation. Accordingly, we define the evolutionary time as  $t_\mathrm{evol} = t - t_0$, where $t_\mathrm{evol} = 0$ marks the beginning of the zoom-in procedure, with the total simulation time given by $t$. For more details on the zoom-in process, see \citet{seifried_2017}.

In total, we consider four simulated MCs, two without magnetic fields named MC1-HD and MC2-HD and two with magnetic fields (see Sect.~\ref{subsec:numerics}), named MC1-MHD and MC2-MHD. The HD simulations are initiated at $t_0 = 11.9~\mathrm{Myr}$, whereas the MHD simulations start at $t_0 = 16~\mathrm{Myr}$. 

\begin{figure}
    \centering
    \includegraphics[width=1\linewidth]{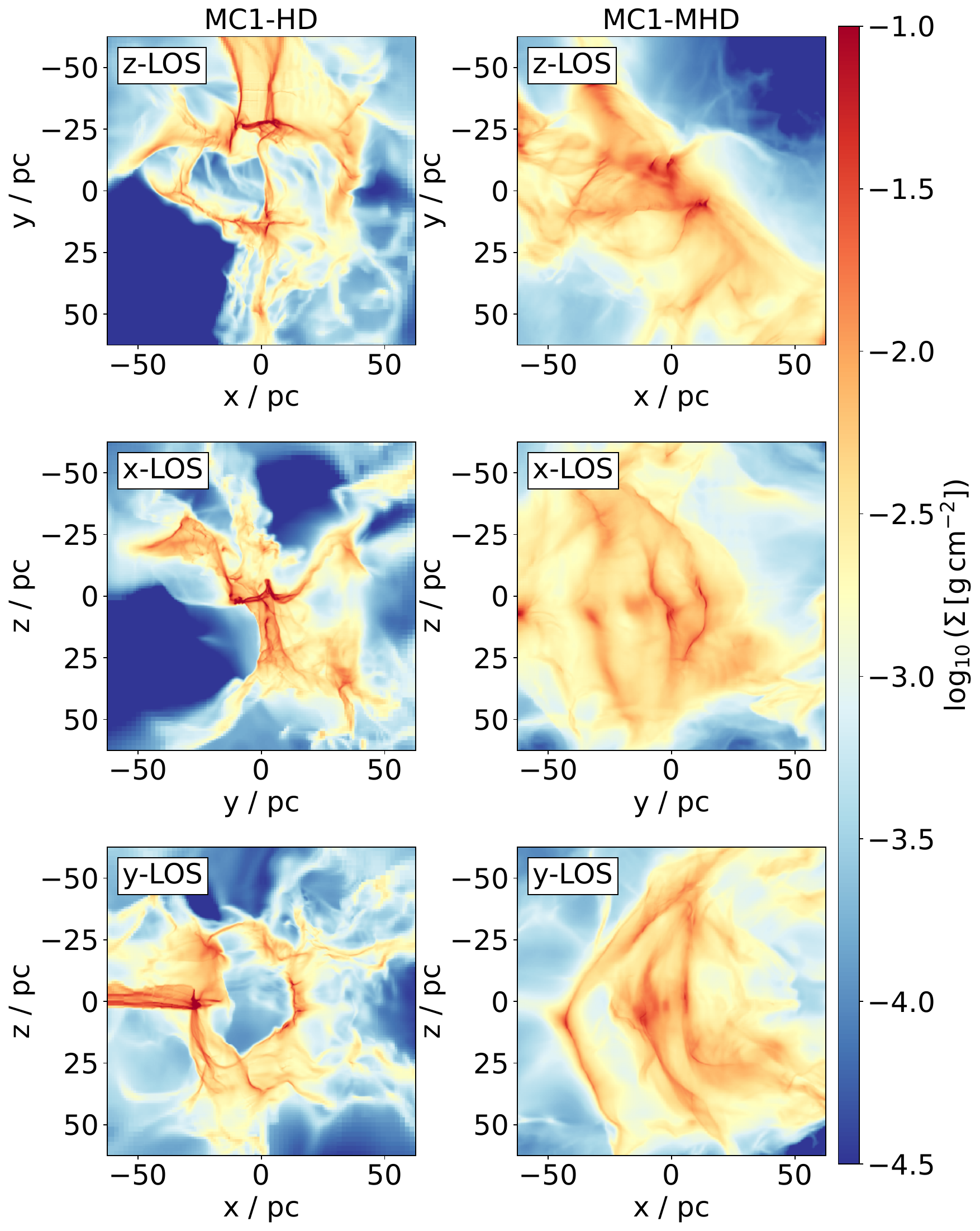}
    \caption{Surface density projections of MC1-HD (\textit{left column}) and MC1-MHD (\textit{right column}) at $t_\mathrm{evol}$ = 2.5~Myr for different projection directions (\textit{top to bottom}). The maps show the complex and filamentary substructure of the clouds. MC1-MHD appears to be more diffuse due to the stabilizing effect of the magnetic field. The color scale is truncated at $\Sigma = 10^{-4.5}~\mathrm{g\,cm^{-2}}$, as the more diffuse material is not relevant for the identification of molecular filamentary structures (Sect.~\ref{subsec:data_prep}), and saturated at $\Sigma = 10^{-1}~\mathrm{g\,cm^{-2}}$ (affecting less than 0.1\% of the pixels) in order to bring out the dense filamentary structure.}
    \label{fig:mcs}
\end{figure}

For the following analysis of the filamentary structure, each cloud is projected along three orthogonal ($x$, $y$, and $z$) lines of sight (LOS), generating the corresponding maps of the gas surface density $\Sigma(x,y)$ (in units of g\,cm$^{-2}$). Since $\Sigma$ corresponds one-to-one to the column density of hydrogen nuclei, $N_\mathrm{H, tot}$ (see Sect.~\ref{subsec:physical_properties}), we follow common usage and also refer to the projected maps as column density maps; the symbol $\Sigma$, however, always denotes the surface density. Each map covers a physical region of 125~pc on a side, resolved at 0.122~pc per pixel (i.e., the same resolution as the underlying simulation), corresponding to a grid of 1024$\times$1024 pixels\footnote{For regions where the AMR grid is more coarsely resolved than the 2D map, we split up the mass of the (3D) cell into the exact proportional masses for the individual (2D) pixels. This procedure does not require any interpolation and thus strictly conserves mass.}.  Although our simulations provide the full 3D structure, we deliberately analyze 2D projected maps so our measurements can be compared directly with the observed, projected $L-M$ relation.
We present the column density maps of MC1-HD and MC1-MHD along each LOS at $t_\mathrm{evol}=2.5~\mathrm{Myr}$ in Fig.~\ref{fig:mcs}. The left column shows the HD clouds while the right column contains the MHD clouds. As discussed in \citet{Seifried2020b}, the presence of a magnetic field makes the clouds appear more diffuse, however, in either case clear filamentary structures are recognizable spanning a wide range of column densities.

\section{Filament identification and characterization} \label{sec:methodology}
In this section we outline the methodologies employed for identifying, isolating and characterizing filamentary structures within the targeted MCs. We start with describing the RHT. Next, we describe the algorithm underlying the dendrogram analysis. Finally, we describe the methods used to calculate the physical properties of the filamentary structures. 

\subsection{Rolling Hough Transform (RHT)}\label{subsec:RHT}
To identify filamentary structures in the column density maps of the SILCC-Zoom simulations, we applied the RHT \citep{Clark_2014}. The RHT is based on the classical Hough Transform \citep{hough_1962}, designed to highlight coherent linear features without requiring sharply defined edges. This makes it particularly suitable for detecting diffuse, interconnected filamentary structures in MCs as well (see Fig.~\ref{fig:mcs}).

The method first enhances small-scale features through high-pass filtering of the surface density map $\Sigma(x,y)$ with an unsharp mask, following \citet{Clark_2014}: $\Sigma$ is smoothed with a circular top-hat kernel whose diameter ($D_K$) is set by the smoothing radius \textit{smr}, and the smoothed map is subtracted from the original to yield the filtered map $\Sigma_\mathrm{US}(x,y)$. The radius \textit{smr} therefore sets the spatial scale of the large-scale structure that is suppressed.
A binary bitmask $B(x,y)$ is then generated by thresholding $\Sigma_\mathrm{US}(x,y)$:
\begin{equation}
    B(x,y) = 
\begin{cases}
1, & \text{if } \Sigma_{\text{US}}(x,y) > 0, \\
0, & \text{otherwise}.
\end{cases}
\end{equation}
Within a local circular window of diameter $D_W$ centered on each pixel, the algorithm evaluates the presence of linear structures by scanning over orientations $\theta$. The traditional Hough transform relation, $\rho=x\cdot \textrm{cos}\theta+y\cdot \textrm{sin}\theta$, is simplified by setting $\rho$=0, reducing the transform to depend only on the orientation angle. For each $\theta$ the RHT measures the fraction of aligned pixels within the local window. If this fraction exceeds $Z$, the corresponding value (i.e., the fraction itself) is recorded for that pixel at that orientation. In this way, the algorithm builds an orientation distribution $R(\theta,x,y)$ at each pixel, quantifying the strength of the linear alignment as a function of angle. Summing over all angles gives a backprojected map,
\begin{equation}
R(x, y)=\int R(\theta, x, y) \, d\theta \, ,
\label{eq:R}
\end{equation}
which quantifies the total linear coherence at every pixel.  Following \citet{erceg_2024}, the normalized backprojection $R(x,y)$, which ranges between 0 and 1, can be interpreted as the likelihood that a given pixel belongs to a coherent linear structure. 

For this work, we used the publicly available RHT code\footnote{\url{https://github.com/seclark/RHT}}. For more details on the method, refer to \citet{Clark_2014}. The algorithm depends on three key parameters:
\begin{itemize}
    \item \textit{smr}: Radius of the unsharp-mask top-hat kernel (related to $D_K$ in \citealt{Clark_2014}); it sets the scale of the large-scale structure suppressed by the high-pass filtering. 
    \item \textit{wlen}: Diameter $D_W$ of the circular window for detecting linear features.
    \item \textit{frac (Z\footnote{$Z$ corresponds to the code parameter \textit{frac}.})}: Minimum fraction of aligned pixels needed to classify a feature as coherent.
\end{itemize}

We optimized these parameters through extensive testing and visual inspection to ensure robust filament identification over the full column density range, resulting in values of 9, 7, and 0.7 for \textit{smr}, \textit{wlen} and \textit{frac}, respectively. Examples of the RHT output using different parameter combinations as well as the motivation for our choice of parameters are provided in Appendix~\ref{subapp:opt_rht}.

\subsection{Dendrogram analysis}\label{subsec:dendrogram}
To isolate individual filamentary structures in the RHT output maps, we employed the dendrogram analysis using the open-source Python package \texttt{AstroDendro}\footnote{\url{https://dendrograms.readthedocs.io}} \citep{astrodendro_2019}. Unlike the more common approach of constructing dendrograms directly from the column density maps, we applied the analysis on the RHT output. This choice allows us to trace the full hierarchical structure of filaments over a wide range of column density values. 

Dendrograms are hierarchical tree diagrams that represent how data cluster together at different levels \citep{rosolowsky_2008}, making them a powerful tool for separating overlapping or nested features in astronomical images.
In this work, dendrograms group together connected pixels according to the output of the RHT algorithm, i.e., $R(x,y)$ (Eq.~\ref{eq:R}), which quantifies the degree of local linear coherence. The individual structures are identified as branches (structures that split into substructures) and leaves (structures without substructure). For a visual definition of these structures, we refer to Fig.~\ref{fig:definition}. This hierarchical separation allows us to extract individual filamentary structures from the broader filamentary network.

The dendrogram construction is controlled by three key parameters (see also the documentation for more details):
\begin{itemize}
    \item \textit{min\_value}: Sets the minimum value of $R$ above which the dendrogram routine is applied, thus filtering out noise.
    \item \textit{min\_delta}: Defines the minimum height in $R$ of a structure (i.e., the difference between its edge and its peak value); decides whether a structure is considered separately or merged with its surrounding structure.
    \item \textit{min\_npix}: Specifies the minimum number of pixels a structure must contain to be recognized as an independent entity.
\end{itemize}
The parameter selection process is described in detail in Appendix~\ref{subapp:opt_astrodendro}. We chose 0.7, 0, and 10 for \textit{min\_value}, \textit{min\_delta}, and \textit{min\_npix}, respectively. This ensures that the extracted structures are physically meaningful while minimizing the inclusion of artifacts.  Because $R(x,y)$ encodes a per-pixel likelihood for linear-coherence rather than a physical intensity, \textit{min\_delta} has no direct physical meaning here; we therefore set it to zero, leaving the separation of nested structures to the physical size criterion \textit{min\_npix}. With a pixel scale of 0.12~pc and \textit{min\_npix}~=~10, this imposes a lower limit of $\approx0.14~\mathrm{pc^2}$ in projected area on any reported substructure (i.e., a linear scale of a few tenths of a parsec). Finally, once the dendrogram is constructed, it is traversed to extract individual filamentary structures for further physical characterization. This provides us with a binary mask indicating whether a pixel belongs to a filament or not. In summary, this method effectively isolates the linear features highlighted by the RHT, enabling robust measurements of their morphological and physical properties.

\subsection{Physical properties of identified filaments} \label{subsec:physical_properties}
Once individual filamentary structures are isolated through dendrogram analysis, we compute their basic physical properties: mean column density, mass, length, width and aspect ratio as well as the critical and virial line mass. In the following we briefly describe how each of them is calculated.

\subsubsection{Mass and mean column density}
\label{sec:mass}
\paragraph{Surface density:} For each identified filament, we compute the mean surface density as the average of the surface density values of all pixels belonging to that structure:
\begin{equation}
    \left\langle \Sigma \right\rangle = \frac{1}{N_{\mathrm{pix}}} \sum_{i=1}^{N_{\mathrm{pix}}} \Sigma_i \, ,
\end{equation}
where $\Sigma_i$ is the surface density value at pixel $i$ and $N_\mathrm{pix}$ is the total number of pixels in the structure. We convert the mean surface density (in units of g~cm$^{-2}$) into a mean particle column density of hydrogen nuclei (units of $\mathrm{cm^{-2}}$) by
\begin{equation}
    \left\langle N_{\mathrm{H,tot}} \right\rangle = \frac{\left\langle \Sigma \right\rangle}{1.4\,m_p} \,
,\end{equation}
where $m_p$=1.67$\times 10^{-24}$~g is the proton mass and the factor 1.4 accounts for the contribution of helium in the ISM\footnote{Note that $N_{\mathrm{H,tot}}$ takes into account all hydrogen nuclei independent of whether they are bound in H$_2$ (thus accounting for two nuclei), H, or  H$^+$ and should not be mistaken for the atomic hydrogen column density. The factor of 1.4 arises from the fact that in the simulations we assume that per hydrogen nucleus 0.1 helium atoms are present, in agreement with primordial abundances.}. 

\paragraph{Mass:} The mass of a filament is determined by integrating its mean surface density over the projected area:
\begin{equation}
    M = \left\langle \Sigma \right\rangle \times A_\mathrm{pix} \times N_\mathrm{pix} \, ,
\end{equation}
where $A_\mathrm{pix}$ is the physical area corresponding to a single pixel, based on the simulation resolution and scaling, and $N_\mathrm{pix}$ is the number of pixels contained in the filament.

We note that $M$ is computed from the raw surface density within the projected footprint of each structure, i.e., without subtracting a local background, which is deliberate.    
Although the dendrogram selects individual filamentary structure in the plane of the sky, from an observational point of view a comparable separation along the LOS is not possible: the column density measured within a structure's footprint cannot be uniquely split into a contribution of the structure itself and of the material lying in front of and behind it. 
Moreover, in a hierarchically structured cloud the latter is to a large extent the structure's own parent, so subtracting a background implies an assumption about where one level of hierarchy is taken to end and the next to begin.
We therefore directly quote the total mass within the projected footprint, keeping in mind that it can also contain a contribution from the larger-scale environment.
In Appendix~\ref{app:background} we nevertheless quantify the impact of a local background subtraction: while the absolute masses are reduced significantly, the $L-M$ relation remains clearly sub-linear ($\langle\alpha\rangle \simeq 0.34$), so none of the conclusions of this work depends on this choice.

\subsubsection{Critical and virial line mass}
\paragraph{Critical line mass ($m_\mathrm{crit}$):} This is the maximum line mass above which an isothermal, cylindrical filament is gravitationally unstable and can collapse to form stars \citep{Stodolkiewicz_1963, ostriker_1964, inutsuka_1992, Inutsuka_1997}:
\begin{equation}\label{eq:m_crit}
    m_\mathrm{crit}(T) = \frac{2c_s^2}{G} \simeq 16.6 \left(\frac{T}{10~ \mathrm{K}}\right)~\mathrm{M_\odot~pc^{-1}} \, ,
\end{equation}
where $c_s$ is the isothermal sound speed, $G$ the gravitational constant, and $T$ the temperature.
Since this estimate assumes isothermality, cylindrical symmetry, isolation, and the absence of magnetic fields  -- conditions rarely satisfied in the ISM -- many observed filaments exceed their critical value \mbox{\citep{hacar_2023}}.

\paragraph{Virial line mass ($m_\mathrm{vir}$):} Accounting for non-thermal motions yields the virial line mass
\begin{equation}
  m_\mathrm{vir} = 2\sigma_\mathrm{tot}^2/G \, ,
\end{equation} where $\sigma_\mathrm{tot}$ is the total velocity dispersion \mbox{\citep[see e.g.,][]{fiege_2000, Li_2016, hacar_2023}}. Adopting the empirical size--linewidth relation compiled by \mbox{\citet{hacar_2023}}, $\sigma_\mathrm{tot}/c_s = \left(1+L/0.5~\mathrm{pc}\right)^{0.5}$, this becomes
\begin{equation}
\label{eq:m_vir}
    m_\mathrm{vir} \simeq \frac{2c_s^2}{G}\left(1+\frac{L}{0.5~\mathrm{pc}}\right) \, ,
\end{equation}
i.e., the line mass at which the filament is in virial balance, and
which is larger than $m_\mathrm{crit}$.

\subsubsection{Length and aspect ratio} \label{subsubsec:length_and_ar}
Since filamentary structures have rather complex shapes, it is difficult to accurately calculate their end-to-end length. We employ two complementary approaches: skeletonization, providing the length estimate $L_{\mathrm{skel}}$ and the minimum enclosing circle (MEC) method, providing the length estimate $L_{\mathrm{MEC}}$.

\begin{figure}
    \centering
    \includegraphics[width=1\linewidth]{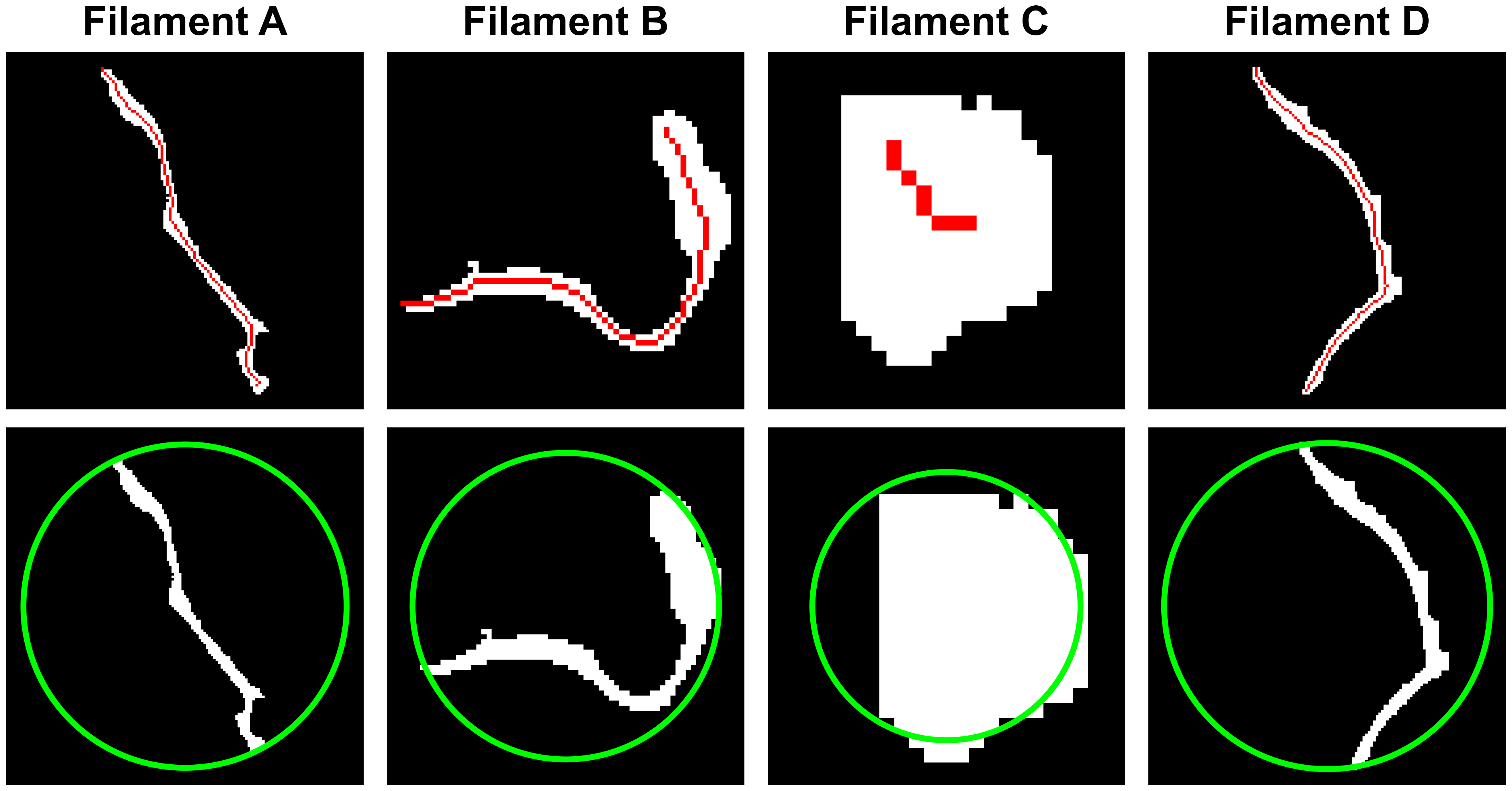}
    \caption{Comparison of two complementary methods for estimating filament geometry. \textit{Top}: Illustration of skeletons obtained by the Zhang -- Suen thinning algorithm. The filaments are depicted in white and their skeletons are shown in red. \textit{Bottom}: MEC obtained with \texttt{OpenCV} for the four filaments. The MEC is shown in green and the filament in white. The figures are not to scale; zoom levels vary to highlight details. Both the methods are complementary to each other, the skeleton gives a good estimate of elongated and highly curved filaments while the MEC diameter is a good estimate for roundish filaments.}
    \label{fig:skl_mec_examples}
\end{figure}

\paragraph{Skeletonization ($L_{\mathrm{skel}}$):}\label{par:skeletonize} We used the Zhang -- Suen thinning algorithm \citep{zhang_1984} to reduce each filament’s binary mask, obtained with the dendrogram algorithm (see Sect.~\ref{subsec:dendrogram}), to a one-pixel-wide skeleton while preserving its topological structure. The binary mask is obtained using the \texttt{get\_mask} method from the \texttt{AstroDendro} package, where pixels belonging to the filament are assigned a value of 1. Skeletonization of the binary mask is then performed using the implementation provided in the \textit{scikit-image} library (\texttt{skimage.morphology.skeletonize}\footnote{\url{https://scikit-image.org/docs/0.25.x/auto_examples/edges/plot_skeleton.html}}). The skeletons for a few selected binary masks are illustrated in the top panel of Fig.~\ref{fig:skl_mec_examples}. It can be seen that the skeletonization method provides a good approximation of the filament length for elongated, linear structures (top panel, filament A, B and D). However, it tends to underestimate the length for more roundish or irregularly shaped features (top panel, filament C), where -- in case of a perfect circle -- the skeleton may collapse even into a single point due to the absence of a well-defined elongation axis. We denote the skeleton length as $L_{\mathrm{skel}}$.

\paragraph{Minimum enclosing circle ($L_{\mathrm{MEC}}$):}\label{par:MEC} To mitigate the aforementioned limitation, we supplement the skeleton-based length with the MEC approach. Using the \texttt{minEnclosingCircle()} function from the \texttt{OpenCV} library, we computed the smallest circle that fully encloses the 2D projection of each filament. The diameter of this circle, $L_{\mathrm{MEC}}$, is then taken as a lower bound on the filament length, providing a direction-independent measure of its spatial extent.

While the MEC systematically underestimates the length of curved linear structures (since it does not account for the curvature of the structure), it remains robust in cases where skeletonization fails, particularly for roundish or irregular morphologies. In this sense, the MEC serves as a lower limit to the filament length. The bottom panel of Fig.~\ref{fig:skl_mec_examples} shows MECs of a few selected filamentary structures, demonstrating this effect. The green circles denote the MECs, and by construction, the length of a filament cannot be smaller than the diameter of its corresponding MEC.

Hence, in order to determine the length of each filament, we compare the length estimates from both methods and take the maximum of the two as the filament length:
\begin{equation}
    L = \mathrm{max}(L_\mathrm{skel}, L_\mathrm{MEC}) \, .
\end{equation}
This ensures that both curved and roundish structures are adequately represented, while minimizing underestimation biases inherent to either method.

\paragraph{Aspect Ratio:}\label{par:AR} The aspect ratio ($AR$) of a filament is defined as the ratio of its length to its width. The width, which in this work is only required to compute the aspect ratio used to select elongated structures (Sect.~\ref{subsec:data_prep}), is measured with an image-processing technique known as the distance transform. Since the technical details of this measurement are not essential for the main analysis, we describe them in Appendix~\ref{app:width}.

\section{Results: The ensemble of filaments} \label{sec:results}

\subsection{Data preparation and parameter optimization}\label{subsec:data_prep}
We analyzed the column density maps of all four clouds --- MC1-HD, MC2-HD, MC1-MHD, and MC2-MHD --- projected along the $x-, y-$ and $z$--direction at evolutionary times $t_\mathrm{evol}$=1.5, 2.5, and 3.5~Myr. For brevity, we present in the following the full analysis, as outlined in Sect.~\ref{sec:methodology}, for the column density map of MC1-HD at $t_\mathrm{evol}$=2.5~Myr projected along the $z$-direction only. We emphasize that the results for the other evolutionary times, clouds, and LOS are qualitatively and quantitatively consistent with those shown here.

As we are interested in assessing the impact of resolution on the identification and characterization of filaments, we consider column density maps at resolutions differing by a factor of 2 in pixel size. For this purpose we consider maps with pixel sizes of $\Delta$x = 0.12~pc, 0.24~pc, 0.48~pc and 0.96~pc.

The final parameters for the RHT and dendrogram analysis are summarized in Table~\ref{tab:parameters}. As noted before, details of the determination of optimal parameter settings for both algorithms at the highest resolution are given in Appendix~\ref{app:para_opt}. However, since the appearance of structures changes with resolution, leading in turn to different RHT outputs, we adapted the \textit{min\_value} parameter in \texttt{AstroDendro} at each resolution. This ensures that the set of identified filamentary structures remained comparable to the highest-resolution case. The specific choice of this parameter at each resolution is detailed in Appendix~\ref{app:choice_of_min_val}.

\begin{table}
\caption{Parameters adopted for the RHT and the dendrogram analysis.}
\label{tab:parameters}
\centering
\begin{tabular}{cc}
\toprule
\multicolumn{2}{c}{RHT} \\
\midrule
\texttt{wlen} & 7 \\
\texttt{smr}  & 9 \\
\texttt{frac} & 0.7 \\
\bottomrule
\end{tabular}

\vspace{0.5em}

\begin{tabular}{ccccc}
\toprule
\multicolumn{5}{c}{\texttt{AstroDendro}} \\
\cmidrule(lr){1-5}
$\Delta$x [pc]& 0.12 & 0.24  & 0.48  & 0.96  \\
\midrule
\texttt{min\_value} & 0.7 & 0.59 & 0.44 & 0.32  \\
\texttt{min\_delta} & 0.0 & 0.0 & 0.0 & 0.0  \\
\texttt{min\_npix} & 10 & 10 & 10 & 10  \\
\bottomrule
\end{tabular}

\tablefoot{The RHT parameters are identical for all resolutions. For the dendrogram, the \texttt{min\_value} parameter was adapted to each resolution, while \texttt{min\_delta} and \texttt{min\_npix} were kept fixed (see Appendix~\ref{app:choice_of_min_val}).}
\end{table}

\begin{figure}
\centering
\hspace{-0.04\linewidth}
\subfigure
{\includegraphics[width=0.46\linewidth]{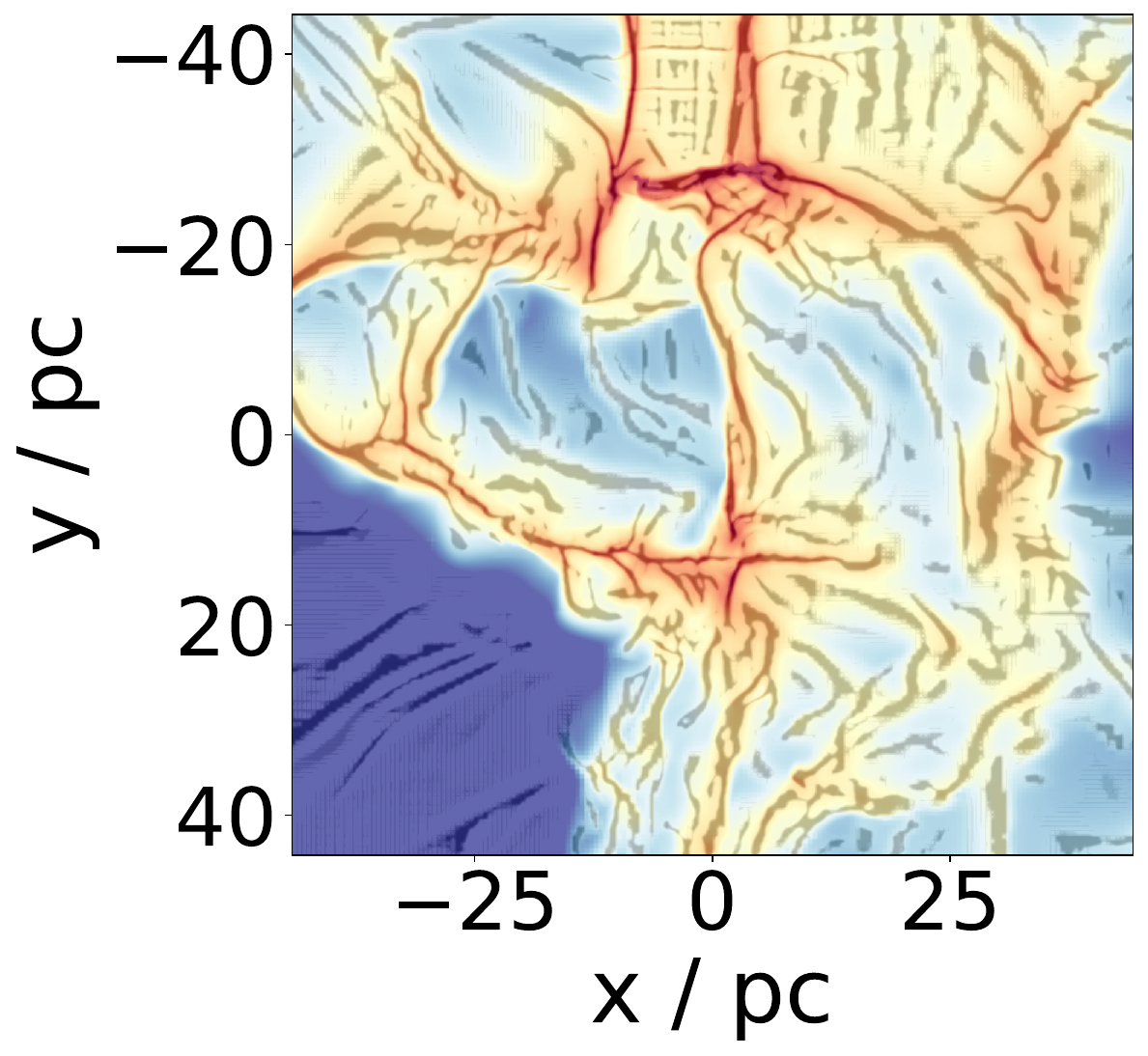}}
\subfigure
{\includegraphics[width=0.46\linewidth]{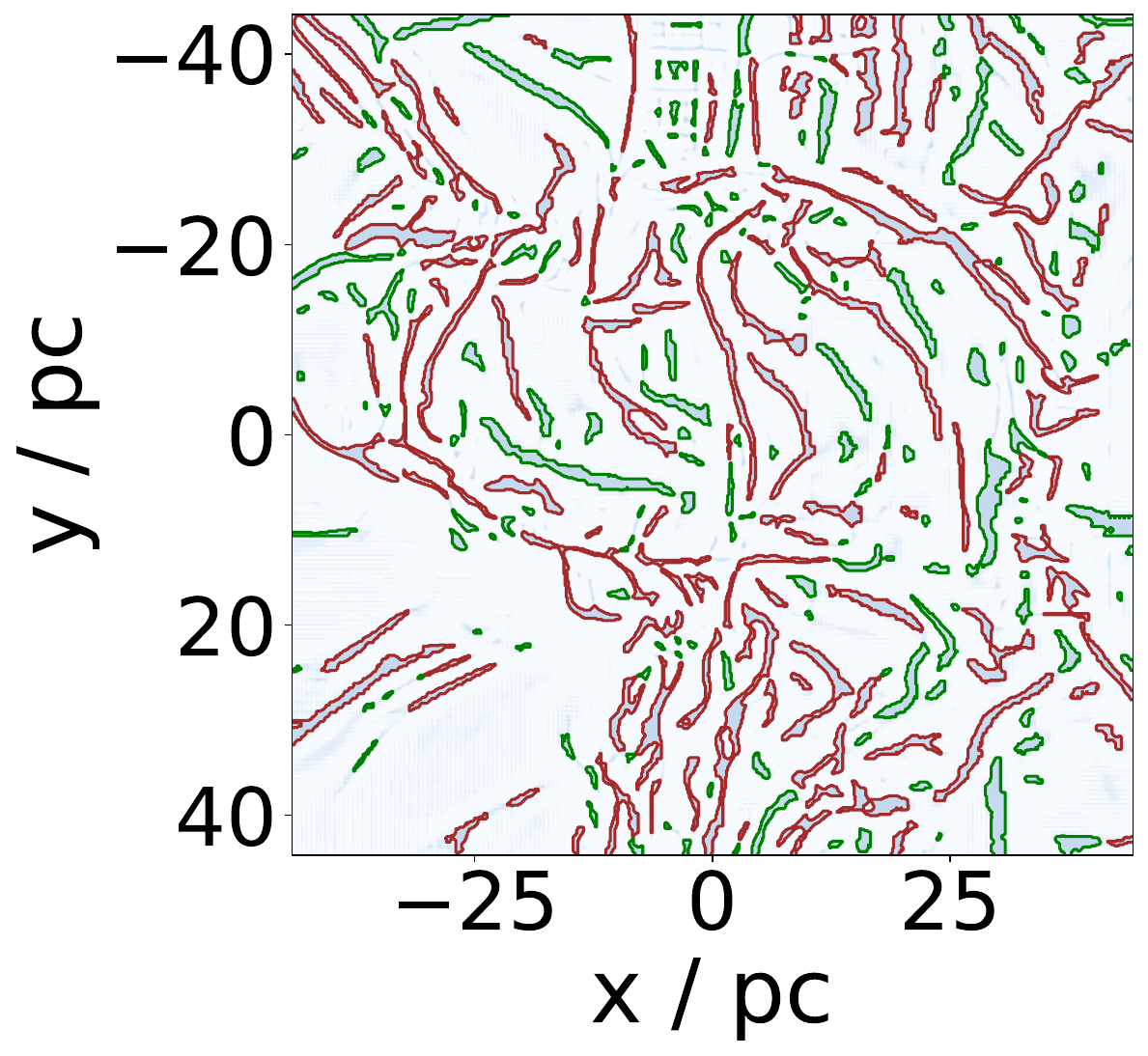}}
\caption{\textit{Left}: RHT output of the column density map for the $z$-projection of MC1-HD at $t_\mathrm{evol}=2.5~\mathrm{Myr}$, overlaid on the corresponding column density map. The map displays the artifact-free central zoom-in region ($\sim$ 87~pc).
\textit{Right}: Filamentary structures identified through the dendrogram analysis of the RHT output. The hierarchical structure is illustrated with base level branches shown in brown and base level leaves in green.}
\label{fig:rht_and_dendrogram}
\end{figure}

To ensure the reliability of our analysis, we focus on the central region of the column density map (approximately 43.5~pc from the center on each side), where grid artifacts are minimal. These artifacts are typically introduced by the transition between the high-resolution zoom-in region and the surrounding lower-resolution volume, an inherent feature of zoom-in simulations. They occur in areas of column density maps, which have a higher resolution than the underlying 3D grid, i.e., particularly near the boundaries of the zoom-in regions (e.g., see lower left corner in Fig.~\ref{fig:wlen_7}). By removing these boundary regions (typically 9~pc), we obtain a clean, almost artifact-free region spanning approximately 87~pc across. The left panel of Fig.~\ref{fig:rht_and_dendrogram} presents the RHT output overlaid on this refined column density map, highlighting the filamentary structure with improved clarity and minimal contamination from boundary effects.

Subsequently, the dendrogram analysis is performed on this cropped RHT output to identify individual filamentary structures. The resulting decomposition distinguishes filamentary structures as branches and leaves within the dendrogram structure. 
Since we focus in this study on the filamentary structure within MCs, we only consider structures with a mean column density $\left\langle N \right\rangle$ > $10^{21}~\mathrm{cm^{-2}}$ and an aspect ratio greater than 3. These criteria further minimize artifacts and simultaneously constrain the selection to molecular filamentary structures, thereby ensuring that only well-defined elongated structures enter the subsequent characterization\footnote{If these criteria are not applied, the dendrogram also identifies hubs and non-molecular filaments. Since these structures are not the focus of this study, we exclude them from our dataset.}. The ensemble of these filaments identified for a given column density map accounts for approximately 20-30\% of the molecular gas mass in the cloud. We note, however, that due to the finite spatial resolution of our simulations, we are not able to resolve sub-filamentary structures such as fibers.

For demonstrative purposes, in the right panel of Fig.~\ref{fig:rht_and_dendrogram}, only the base-level structures are shown (see Fig.~\ref{fig:definition} for the definition of base-level structures), color-coded according to their classification (branches in brown and leaves in green). Most of the identified structures at this level correspond to branches, though a few isolated leaves without substructure are also visible, providing a clear view of how the filamentary network is segmented. We also emphasize that filamentary structures of various morphologies are present -- from linear to twisted and branched. 

\subsection{The \texorpdfstring{$L-M$}{L-M} relation: Setting the stage}\label{subsec:m-l_relation}

After this preparative work, we now turn to one of the central diagnostics of filamentary structures: the $L-M$ relationship, which was systematically compiled and characterized for filamentary structures by \citet{hacar_2023}. This relation is particularly important as it suggests the existence of relations connecting filaments at different scales which may reflect universal physical processes governing their evolution in the ISM.
To test the robustness of these trends, we examine the $L-M$ relation for all simulated MCs at multiple spatial resolutions, assessing whether the observed relation is intrinsic or resolution-driven.

Before doing so, we recapitulate some of the key quantities discussed in \citet{hacar_2023}, which are also considered in our analysis in Sects.~\ref{subsec:filaments_at_multi_res} and~\ref{subsec:hierarchical_decomp}.
Two of these quantities are the critical line mass (Eq.~\ref{eq:m_crit}) and virial line mass (Eq.~\ref{eq:m_vir}) already discussed before. Furthermore, since the observability of filaments via molecular line transitions is closely tied to their ability to shield molecular gas from dissociating ultraviolet (UV) radiation, we also take into account the conditions required for molecular tracers such as \ce{CO} to form and survive. This shielding requires a minimum column density, $N_\mathrm{shield}$, which in turn is set by the filament’s line mass ($m$) and its inner (flat) radius $R_\mathrm{flat}$. Based on this, one can derive an approximate scaling relation between a filament's mass and its length \citep[see Eq. 22 in][]{hacar_2023}:
\begin{equation}\label{eq:shielding_line}
    L \approx 1.5 \left(\frac{M}{\mathrm{M_\odot}}\right)^{0.65}\left(\frac{N_\mathrm{shield}}{10^{21}~\mathrm{cm^{-2}}}\right)^{-0.65}~\mathrm{pc} \, . 
\end{equation}
Eq.~\ref{eq:shielding_line} connects the column density threshold for shielding of UV radiation, $N_\mathrm{shield}$, to the physical size and mass of the filament, ultimately offering insight into the conditions necessary for molecular filaments.

\citet{hacar_2023} find that filaments are located in the $L-M$ plane mostly between the two lines given by Eq.~\ref{eq:shielding_line} and that derived from the virial mass (Eq.~\ref{eq:m_vir}).
Furthermore, the authors show that the distribution of these filaments can be described by a power law of the form 
\begin{equation}
L\propto M^\alpha \, .
\label{eq:M-L}
\end{equation}
In the following, we fit such a power law to our data. In order to ensure a uniform statistical weight across the mass range (i.e., the $x$-axis), we bin the data points logarithmically such that each mass dex contributes equally. A linear regression is then performed on the binned data in logarithmic space, which gives the `slope' $\alpha$.

\subsection{The \texorpdfstring{$L-M$}{L-M} relation across multiple resolutions}\label{subsec:filaments_at_multi_res}

Filamentary structures in MCs exist across a wide range of scales. It is thus necessary to combine different observing facilities with different resolutions to uncover the properties across all involved scales. In consequence, derived properties may depend on the observationally available resolution \citep{hacar_2023}. For this reason, we first examine whether and how the $L-M$ relation of filamentary structures is affected by limited observational resolution. This allows us to uncover how coarse observations may obscure the underlying complexity of filamentary networks, influencing the $L-M$ scaling itself, and, for example, the interpretation of gravitational stability.

We note that \citet{Feng_2024} also performed a resolution test concerning filamentary structures, however, with the focus on the robustness of the temporal evolution of the $L-M$ relation. 
In our case, resolution serves as a direct observational proxy for hierarchy: a structure identified at low resolution is, at higher resolution, resolved into the same nested substructures that the dendrogram exposes at fixed resolution. Demonstrating that both routes yield $\alpha\approx0.5$ is the result we are after.
 
For this purpose, we systematically degrade the resolution of the column density maps, initially resolved at 0.12~pc, down to coarser resolutions of 0.24~pc, 0.48~pc, and 0.96~pc. At each resolution level, we perform a consistent sequence of RHT and dendrogram analysis (see Table~\ref{tab:parameters} for parameter values) to extract the filamentary structures. 

\begin{figure}
    \centering
    \includegraphics[width=1\linewidth]{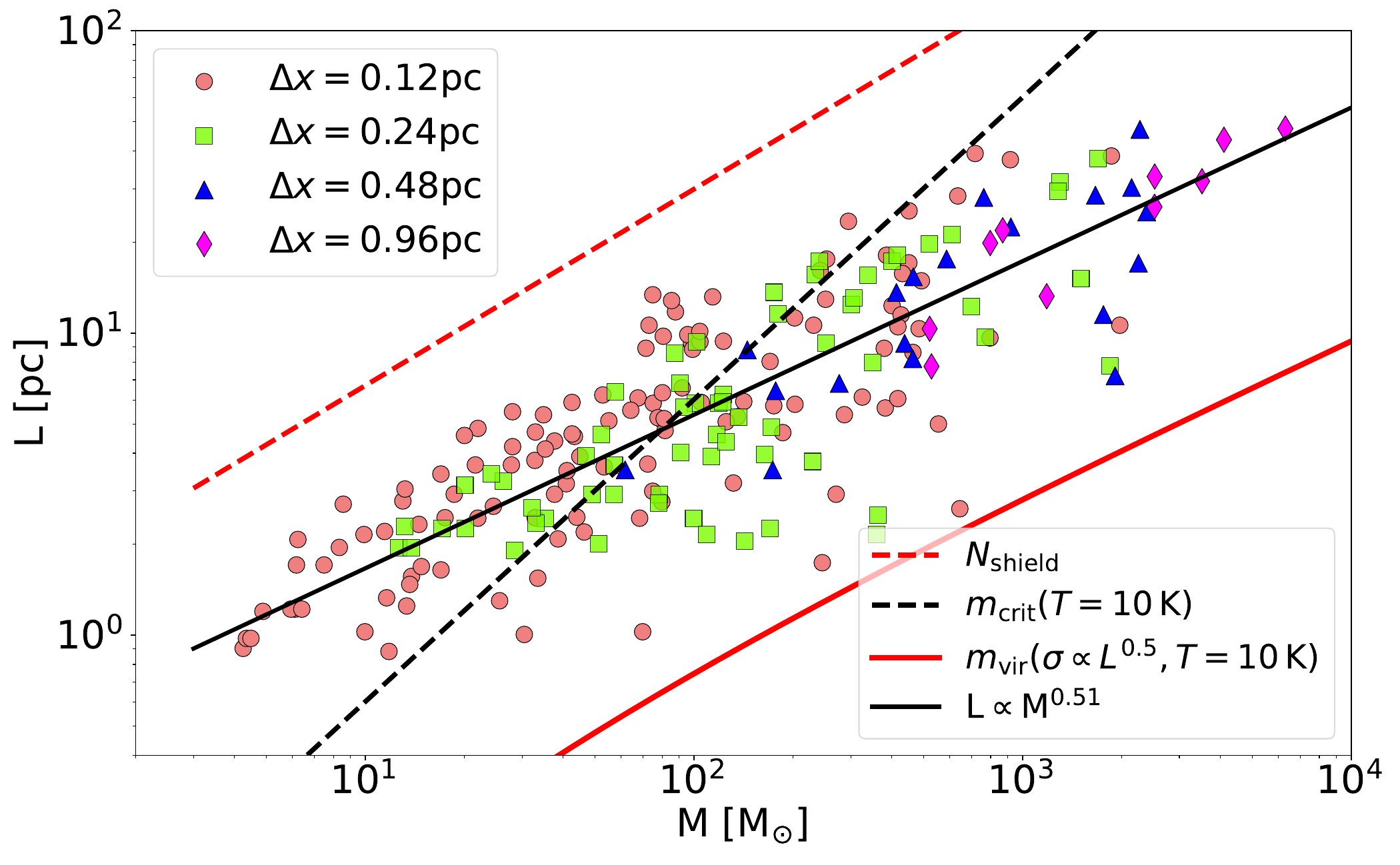}
    \caption{The relation between length and mass for filaments identified at various resolutions (indicated by different colors and marker shapes) for the $z$-projection of MC1-HD at $t_\mathrm{evol}=2.5~\mathrm{Myr}$. The dashed red line corresponds to the shielding threshold defined by Eq.~\ref{eq:shielding_line}, the solid black line shows the power-law fit to the data, the dashed black line marks the critical line mass given by Eq.~\ref{eq:m_crit} for \mbox{$T$ = 10 K}, and the solid red line denotes the virial line mass given by Eq.~\ref{eq:m_vir}. These reference lines are consistently used in all $L-M$ plots presented hereafter. A best-fit power-law index of $\alpha=0.51\pm0.02$ is obtained for our identified filaments, consistent with the findings of \citet{hacar_2023}.}
    \label{fig:M-L_baselevel_all_res}
\end{figure}

The filamentary structures identified across these resolutions, which lie at the base level of the dendrogram (see Appendix~\ref{subapp:opt_astrodendro}, particularly Fig.~\ref{fig:definition} for the definition), are compiled into a single $L-M$ plot, as shown in Fig.~\ref{fig:M-L_baselevel_all_res}. The filaments span a mass range from approximately 1~$\mathrm{M_\odot}$ to a little less than 10$^4$~$\mathrm{M_\odot}$ and a length range from sub-parsec scales up to more than 30~pc, covering approximately three orders of magnitude in both axes. The population of filaments identified in our simulations is located between the column density threshold for molecular gas (Eq.~\ref{eq:shielding_line}; dashed red line) and the equipartition state (derived from Eq.~\ref{eq:m_vir}, red solid line) and thus in agreement with observations \citep{hacar_2023}. It can also be seen that the filaments cover a wide range of line masses (about two~orders of magnitude).

Moreover, the global distribution of the filaments in the $L-M$ phase-space can be described by a power-law relation (Eq.~\ref{eq:M-L}).
When fitting our data in Fig.~\ref{fig:M-L_baselevel_all_res}, we obtain a power-law index of $\alpha = 0.51 \pm 0.02$. Repeating this for all clouds, times and LOS yields a mean power-law slope of $\langle \alpha \rangle = 0.54 \pm 0.07$, indicating a sub-linear scaling between filament mass and length. The derived value is consistent with the result of \citet{hacar_2023}. Notably, \citet{Feng_2024} also reported a similar scaling with $\alpha = 0.45$ for an ensemble of filaments identified across different resolutions.
There is, however, a substantial scatter around the fitted scaling relation. Filamentary structures with similar masses display a range of lengths. This spread reflects the structural diversity among filaments and points to varying formation conditions, evolutionary stages, and environments. 

At higher resolutions, filamentary structures predominantly occupy the lower left part of the $L-M$ range in Fig.~\ref{fig:M-L_baselevel_all_res}; however, as the resolution decreases, a systematic shift toward higher masses and lengths is observed. This trend indicates that filaments identified at lower resolution are often revealed to be composed of multiple, smaller sub-filaments when observed at higher resolution. This behavior is consistent across all clouds, LOS, and evolutionary times.
These findings again match observational results where large-scale filaments observed with single-dish telescopes at low resolution break down into several smaller filaments when observed by interferometers at higher spatial resolution \citep[e.g.,][]{hacar_2018}.

To quantify this, we consider the entirety of the detected structures by combining all clouds, LOS, and evolutionary times. We find that the number of identified base-level  structures drops from more than \mbox{$11 \times 10^{3}$} at \mbox{$\Delta$x = 0.12~pc} to about 700 at 0.96~pc, confirming our interpretation that coarsely resolved, filamentary structures break up into more and smaller structures. Furthermore, in Sect.~\ref{subsec:hierarchical_decomp} we also investigate the hierarchical structure of individual filaments.

\subsection{The \texorpdfstring{$L-M$}{L-M} relation at the highest resolution} \label{subsec:hierarchical_decomp}

\begin{figure}
    \centering
    \includegraphics[width=1\linewidth]{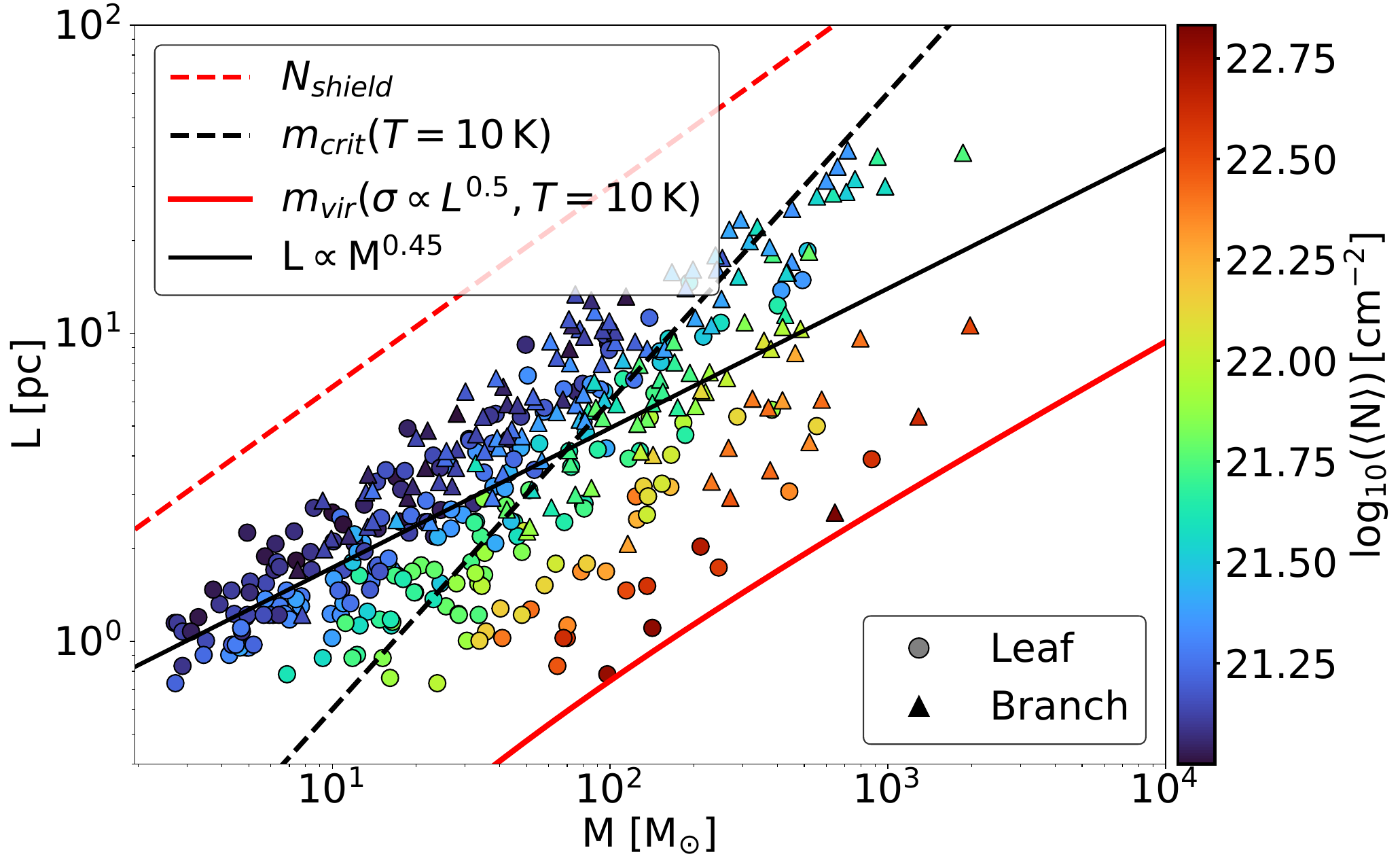}
    \caption{The relation between length and mass for filaments along the $z$-projection of MC1-HD at $t_\textrm{evol} = 2.5~\mathrm{Myr}$, colored by the average column density of each structure. Only filaments identified at the highest resolution are shown. Note that all branch filaments (at each level) are decomposed into their children to create this ensemble. Leaf filaments are shown as circles and branch filaments as triangles. A best-fit power-law index of $\alpha = 0.45 \pm 0.03$ is obtained, similar to that found for the ensemble of filaments at different resolutions in Fig.~\ref{fig:M-L_baselevel_all_res}.}
    \label{fig:M-L_trunk_1024_res_all_structures}
\end{figure}

Building on this baseline, we next examine whether the trend observed in Sect.~\ref{subsec:filaments_at_multi_res} is just a resolution effect or instead reflects an intrinsic hierarchical structure. For this purpose, we decompose branch filaments from the highest resolution column density map (\mbox{$\Delta$x = 0.12~pc}) into their constituent sub-branches and leaves at all dendrogram levels (see Fig.~\ref{fig:definition} for a schematic representation of branches and leaves and also Sect.~\ref{subsubsec:hierarchy_in_ind_fils} for a study of individual branch-leaf-structures). This creates an ensemble of hierarchical structures. By combining all these structures into a unified $L-M$ distribution, we test whether the scaling behavior found in Fig.~\ref{fig:M-L_baselevel_all_res} persists independently of the resolution bias\footnote{We remind the reader that in Fig.~\ref{fig:M-L_baselevel_all_res} we only consider filaments at the base level.}. The result is shown in Fig.~\ref{fig:M-L_trunk_1024_res_all_structures}, where the color scale indicates the average column density. Leaves are represented by dots and branches by triangles. 

Similar to Fig.~\ref{fig:M-L_baselevel_all_res}, the distribution in Fig.~\ref{fig:M-L_trunk_1024_res_all_structures} shows a wide spread around the best-fitting power law, indicating that filaments with similar masses can exhibit a broad range of lengths. Branch-like filaments occupy the upper right portion of the $L-M$ plane, reflecting their larger spatial extent, while shorter filaments populate the lower left region. A detailed examination of the internal variation of $\langle N \rangle$ within this population, including the observed spread at fixed mass, is presented in Sect.~\ref{subsubsec:columndensity}.

We fit a power-law relation (Eq.~\ref{eq:M-L}) to the distribution shown in Fig.~\ref{fig:M-L_trunk_1024_res_all_structures}. We obtain a power-law index of $0.45 \pm 0.03$ for the shown column density map and a mean power-law slope (across all clouds, LOS, and evolutionary times at a resolution of 0.12~pc) of $\langle \alpha \rangle = 0.53 \pm 0.07$, consistent with the mean power-law index found for filaments identified at different resolutions (see Fig.~\ref{fig:M-L_baselevel_all_res}). This reasonable agreement demonstrates, for the first time, that the sub-linear $L-M$ relation -- comprehensively compiled and discussed for filaments by \citet{hacar_2023} -- is an intrinsic property of the filamentary structure of the ISM, rather than an artifact of combining observations with different resolutions.

\subsubsection{Hierarchical structure and depth of filaments}
\label{subsubsec:hierarchical_depth}
The full decomposition allows us to quantify the number of hierarchical levels (i.e., the number of dendrogram levels of sub-branches and leaves below the base level; see Fig.~\ref{fig:definition} for their definition) present in the identified filamentary structures. For each base-level branch, we count the number of dendrogram levels below it: a depth of one corresponds to a base-level branch that splits directly into its leaves, a depth of two to a branch containing one intermediate generation of sub-branches, and so forth. Considering first only the base-level branches (4142 for all clouds, LOS, and evolutionary times at the highest resolution), we find that shallow hierarchies dominate: 45\% of the base-level branches have a depth of one, 24\% a depth of two, 14\% a depth of three, and 17\% a depth of four or more (with a maximum depth of 15), yielding a mean (median) depth of 2.2 (2). Hence, one- and two-level systems together make up about 70\% of the population with substructure, and deeply nested configurations become progressively rarer. When base-level leaves, which have an hierarchical depth of zero, are also included, the average depth drops to about 0.8, since they constitute about 64\% of all base-level structures. We further find that the depth of the hierarchy is almost unaffected by spatial resolution: the mean depth of the branches decreases only marginally, from about 2.2 levels at $\Delta x = 0.12$~pc to 2.0 at 0.96~pc. Including the base-level leaves, the average depth remains around 0.8 for $\Delta x = 0.12$, 0.24, and 0.48~pc, decreasing only to about 0.6 at $\Delta x = 0.96$~pc. Thus, decreasing the spatial resolution primarily reduces the number of detectable structures rather than the hierarchical depth of those that remain identifiable.

\subsubsection{Variation of column density across filaments} \label{subsubsec:columndensity}

In addition to the power-law relation, the filament population shown in Fig.~\ref{fig:M-L_trunk_1024_res_all_structures} exhibits a striking spread in mean column densities, spanning nearly two orders of magnitude,  even at a fixed mass. This dispersion is not random but reflects the intrinsic inhomogeneity of the filamentary network. The large span in column densities is attributed to the ability of RHT (by re-normalizing the column density in each window by its mean) to identify filaments at all scales, irrespective of their column densities. Extended branch-like structures populate the upper region of the $L-M$ plane with comparatively low column densities, consistent with their large spatial extent and inclusion of diffuse material. By contrast, short and compact filaments, typically leaves or embedded sub-filaments, occupy the lower-right region and approach or exceed the critical line mass, indicating a progressively higher impact of gravity. Overall, we find line masses of our filaments (not shown) to span a range from roughly 1 to \mbox{100 M$_\odot$ pc$^{-1}$}.

A systematic trend of increasing column density toward the bottom-right of the diagram suggests that, as filaments evolve or fragment, they accumulate mass into smaller volumes, thereby increasing their density. Physically, this may trace the approach toward gravitational boundedness. The overall two~orders of magnitude in $\langle N \rangle$ are consistent with a hierarchically assembled ISM in which dense leaves coexist within more diffuse parental branches. The relative position of filaments in the $L-M$ plane, depending on their mean column density, has been explored by \citet{hacar_2025}. Using analytic calculations, these authors predict that while filamentary structures should indeed follow a $L = a\cdot M^{\alpha}$ dependence with a slope around \mbox{$\alpha$ = 0.5}, their normalization value $a$ should vary as function of column density: filaments with increasing column density will have lower values of $a$ and are thus located toward the bottom right. Our simulated filaments reproduce this behavior very well exhibiting a similar dependence with column density (see Fig.~\ref{fig:M-L_trunk_1024_res_all_structures}).

\section{Results: Individual filaments}
\label{sec:results_individual}

While the previous analyses rely on ensemble statistics, it is equally important to examine whether the observed relations also hold for individual filamentary structures.
Similar to the previous analyses we investigate this for filaments identified at different resolution as well as at the highest resolution only.

\subsection{The \textit{L -- M} relation for individual filaments across various resolutions} \label{subsubsec:consistency_across_ind_fils}

First, we focus on specific regions of the column density map where we can identify a single filamentary structure at the lowest resolution and then trace the filamentary structure within the same region across varying resolutions. This approach enables us to test whether the observed scaling behavior arises solely from a compilation of different filamentary structures -- or whether it persists when following a single structure consistently across scales.

\begin{figure*}
    \sidecaption
    \includegraphics[width=12cm]{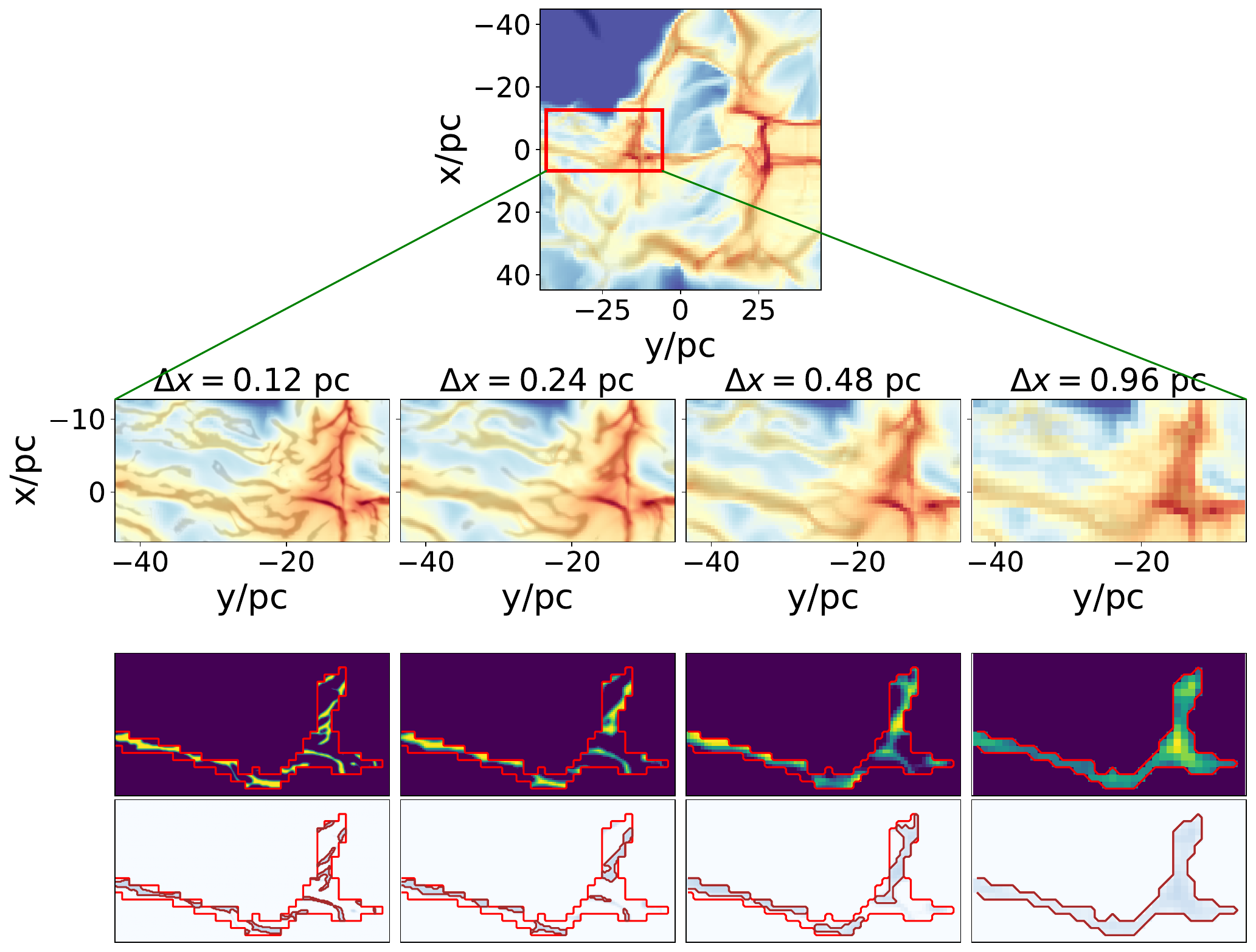}
    \caption{\textit{Top row}: Selected region (highlighted in red) of the column density map along the $z$-direction of MC1-HD at $t_\mathrm{evol}=2.5~\mathrm{Myr}$ -- at the lowest resolution, \mbox{$\Delta$x = 0.96 pc}, with the RHT output overlaid (black shading). \textit{Second row}: A zoomed-in view of the selected region shown at progressively higher resolutions -- \mbox{$\Delta$x = 0.96}, 0.48, 0.24, and 0.12~pc from right to left -- overlaid with the corresponding RHT output (black), which highlights the hierarchy of filamentary structures across scales. \textit{Third row}: Mask (red contour) of the filament selected at a resolution of \mbox{$\Delta$x = 0.96~pc} along with the filamentary structures within it identified at higher resolution. Only the filamentary structures within this region are shown, with all others removed for clarity. The color coding represents the column density of the filamentary structures. \textit{Bottom row}: All filaments detected within the contour region, visualized together to illustrate the complexity and fragmentation of the larger filamentary structure.}
    \label{fig:selected_structure}
\end{figure*}

The step-by-step procedure for this process is illustrated in Fig.~\ref{fig:selected_structure}. We begin by selecting a visually distinguishable filament from the lowest-resolution column density map. A mask delineating the spatial extent of this filament is created from the binary mask obtained via the RHT and dendrogram analysis (red line). This mask is subsequently applied to the higher-resolution maps, thereby ensuring that the same physical region is examined at each level of resolution. Within this masked region, both the RHT and dendrogram analyses are carried out for all higher resolution maps, allowing us to investigate the filament’s structural properties across scales.

Examining the results of this procedure provides important insight into how resolution affects filament identification. The RHT analysis of the selected region (prior to isolating the filament, i.e., the second row of Fig.~\ref{fig:selected_structure}) reveals filamentary structures in the higher-resolution maps at locations where no such features are visible in the coarser resolved column density maps (compare black-shaded regions). This demonstrates that increasing the resolution not only uncovers the hierarchical nature of filamentary structures, where a structure identified at low resolution is resolved into a network of finer substructures, but also reveals an additional population of faint, small-scale filaments that remain hidden at lower resolutions.

\begin{figure}
    \centering
    \includegraphics[width=1\linewidth]{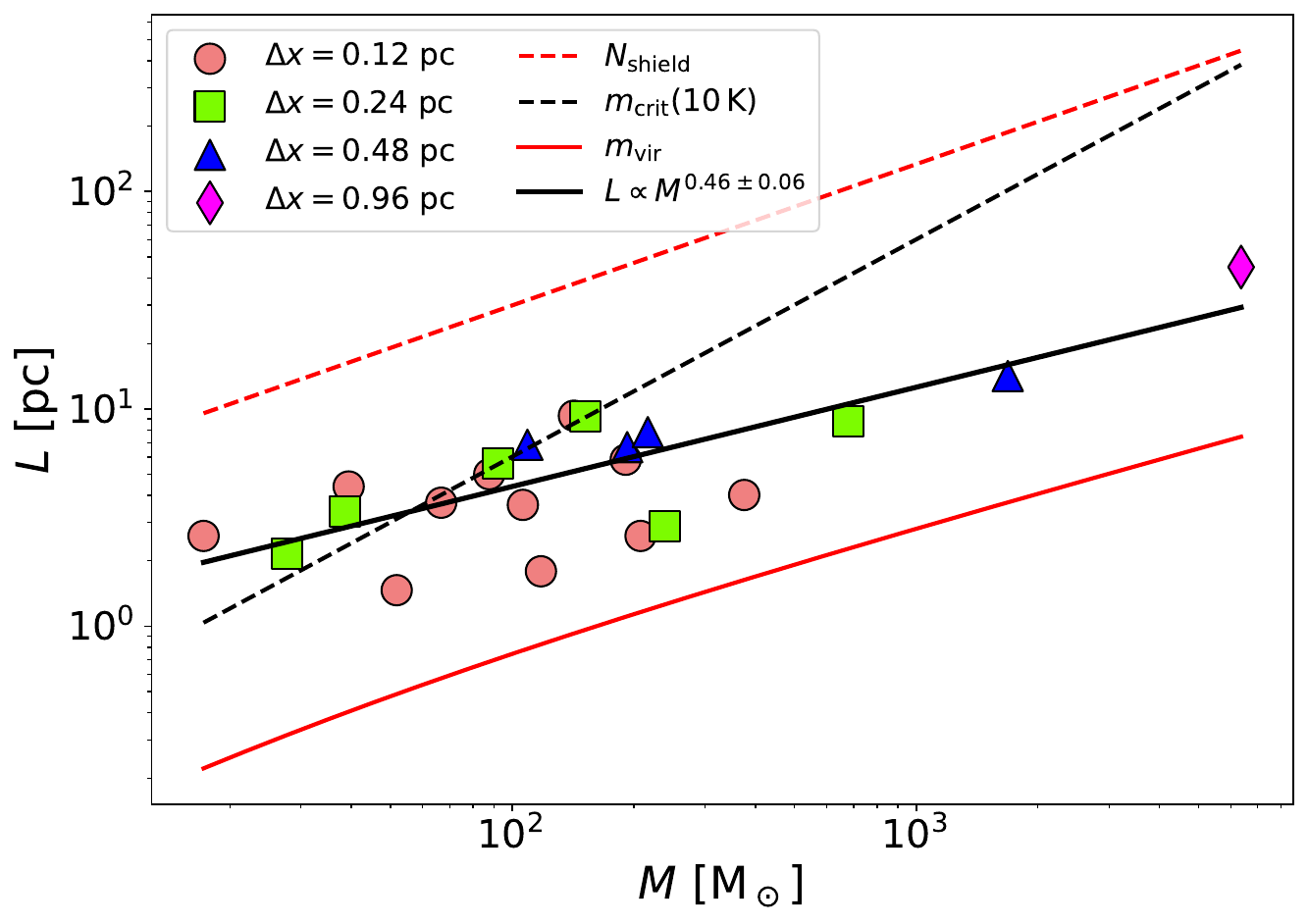}
    \caption{The relation between length and mass for filaments at various resolutions which all are located within the spatial extent of a filamentary structure identified in run MC1-HD at the lowest resolution (see Fig.~\ref{fig:selected_structure}). A sub-linear trend is observed, with a best-fit power-law slope of $\alpha=0.46\pm0.06$.} \label{fig:selected_structure_M_L_plot}
\end{figure}

Building upon this, we then construct a $L-M$ diagram that incorporates all filamentary structures identified within the selected region across the different resolutions in Fig.~\ref{fig:selected_structure_M_L_plot}. Note that we only consider the structures at the base level (see Fig.~\ref{fig:definition} for its definition). For the structure shown here taken from run MC1-HD at \mbox{$t_\textrm{evol} = 2.5~\mathrm{Myr}$}, the resulting distribution exhibits a sub-linear scaling between filament mass and length with a slope of \mbox{$\alpha$ = $0.46\pm0.06$} (Eq.~\ref{eq:M-L}). We further examined several additional structures from the column density maps of different clouds, LOS, and timestamps, and found that the slope consistently remains within the range $\alpha = 0.45 - 0.70$, similar to the previous findings for the ensembles of filamentary structures.

Finally, the $L-M$ relation in Fig.~\ref{fig:selected_structure_M_L_plot} shows that -- as expected -- filaments extracted from coarser column density maps appear systematically longer and more massive than those identified at higher resolutions. However, we  note that in the higher resolution maps, the filaments become highly fragmented and some of these structures fall below the mean particle column density threshold of \mbox{$\left\langle N \right\rangle$ = $10^{21}~\mathrm{cm^{-2}}$} (Sect.~\ref{subsec:data_prep}) and thus are not included in the $L-M$ plot. 

\subsection{The \texorpdfstring{$L-M$}{L-M} relation for individual filaments at the highest resolution} \label{subsubsec:hierarchy_in_ind_fils}

The analysis presented in Sect.~\ref{subsubsec:consistency_across_ind_fils} focused on filamentary structures identified across all resolutions and the aggregated $L-M$ plots were found to exhibit sub-linear trends. To determine whether this behavior arises solely from resolution effects or reflects intrinsic structural properties, we next decompose individual branch filaments identified in the highest-resolution column density map (\mbox{$\Delta$x = 0.12~pc}) into their constituent branches and leaves. By examining the resulting $L-M$ relations of these hierarchical components, we can assess whether the slopes observed in Sect.~\ref{subsubsec:consistency_across_ind_fils} are intrinsic to the filament hierarchy or are imposed by resolution. 

\begin{figure*}
    \centering
    \includegraphics[width=1\linewidth]{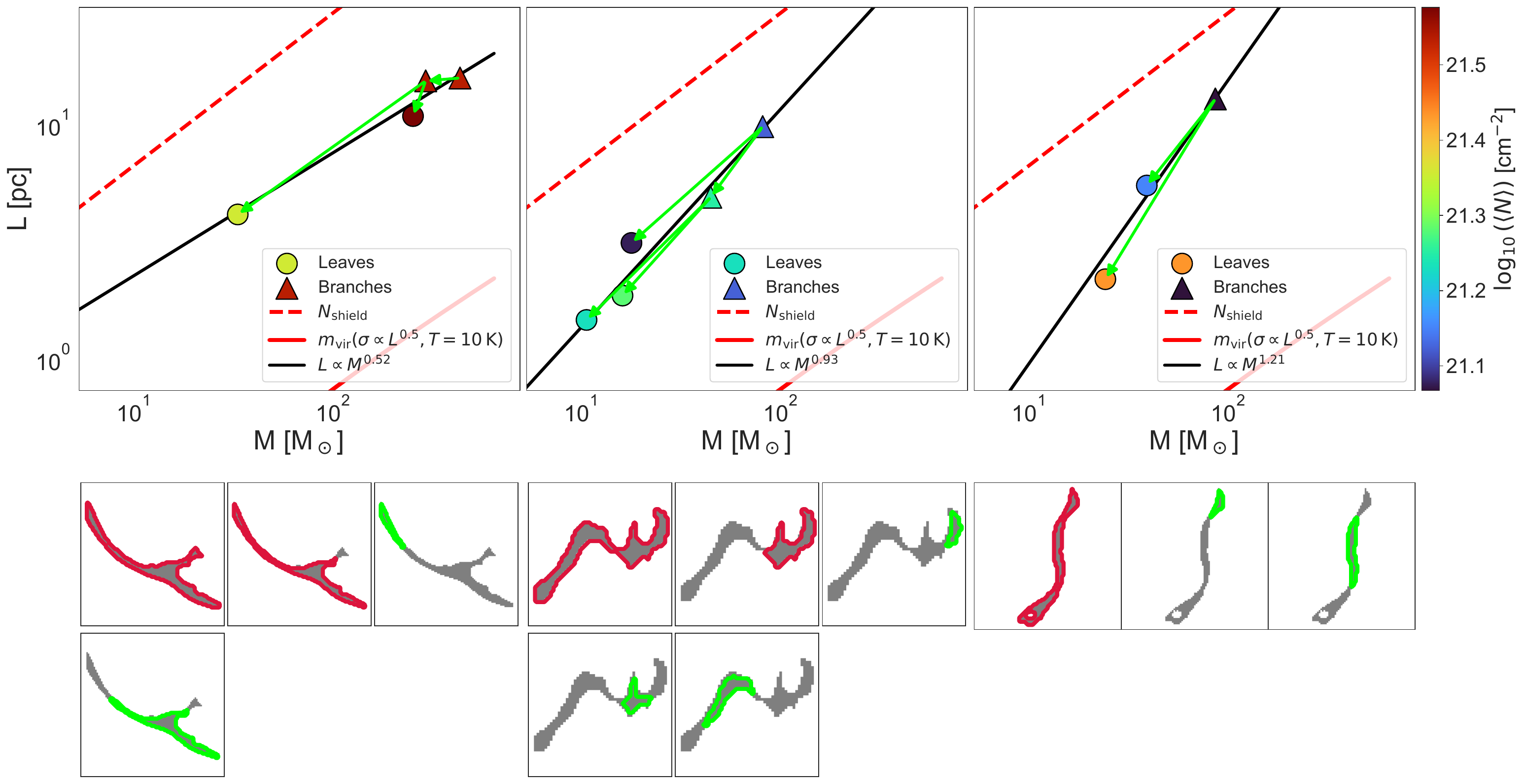}
    \caption{\textit{Top row}: $L-M$ scaling relations for individual, hierarchically segmented filamentary structures. Each panel shows branches and their children, color-coded by mean column density. 
    The solid black line represents a power-law fit of the form $L\propto M^\alpha$, with the best fit given in the legend. The green arrows represent the hierarchy, i.e., how each branch connects to its children. The red lines are the same as in Fig.~\ref{fig:M-L_baselevel_all_res}.
    \textit{Bottom row}: Adjacent to each panel in the top row, we show corresponding maps where the parent filament is displayed in gray and its children are contoured in red (branch) and green (leaf). These examples illustrate how deviations from geometric simplicity and column density homogeneity influence the $L-M$ scaling and produce a wide range of slopes.}
    \label{fig:structures}
\end{figure*} 

\subsubsection{Diversity of \texorpdfstring{$L-M$}{L-M} scaling}
\label{subsubsec:diversity_in_scaling}
For a simple, linear filament with (nearly) uniform column density, one would expect a linear scaling of $L \propto M$ when segmenting the structure longitudinally \citep{hacar_2023}. This clearly deviates from the various scalings shown so far, which all reveal a sub-linear scaling $L \propto M^{\alpha}$ with $\alpha \sim 0.5$. Real filaments, however, are rarely so idealized: departures from a simple cylindrical geometry as well as inhomogeneities in column density both have the potential to alter the scaling.

To illustrate this, we focus on branch filaments identified at the highest resolution in the bottom row of Fig.~\ref{fig:structures}. Each branch is recursively decomposed into its constituent children, and $L-M$ diagrams are constructed for this hierarchy. The resulting relations, along with their fitted power laws for three selected structures, are displayed in top row of Fig.~\ref{fig:structures}. 
Among the three examples, the right panel displays a power-law index greater than unity ($\alpha = 1.27$), the middle panel is consistent with an (almost) linear scaling ($\alpha = 0.93$), and in the left panel we find a sub-linear relation ($\alpha = 0.58$) similar to the previously discussed relations. The outcome of this analysis highlights the diversity of scaling behaviors that can emerge within a single filamentary structure.  We note that projection influences individual structures, though not by biasing $\alpha$ directly: a common inclination foreshortens all lengths by the same factor and therefore shifts the normalization of the $L-M$ relation rather than its slope, while the mass is unaffected. What does alter individual slopes is, first, that filaments can be curved and their substructures may not be aligned with the parent (Fig.~\ref{fig:selected_structure}): hence, the foreshortening differs from sub-filament to sub-filament. Second, the projected mass includes all material along the line of sight, i.e., fore- and background material. Both effects can lead to large variations in the fitted value of $\alpha$ for different structures. The ensemble averages (Sect.~\ref{sec:results}), however, are far more robust.  

\begin{figure}
    \centering
    \includegraphics[width=1\linewidth]{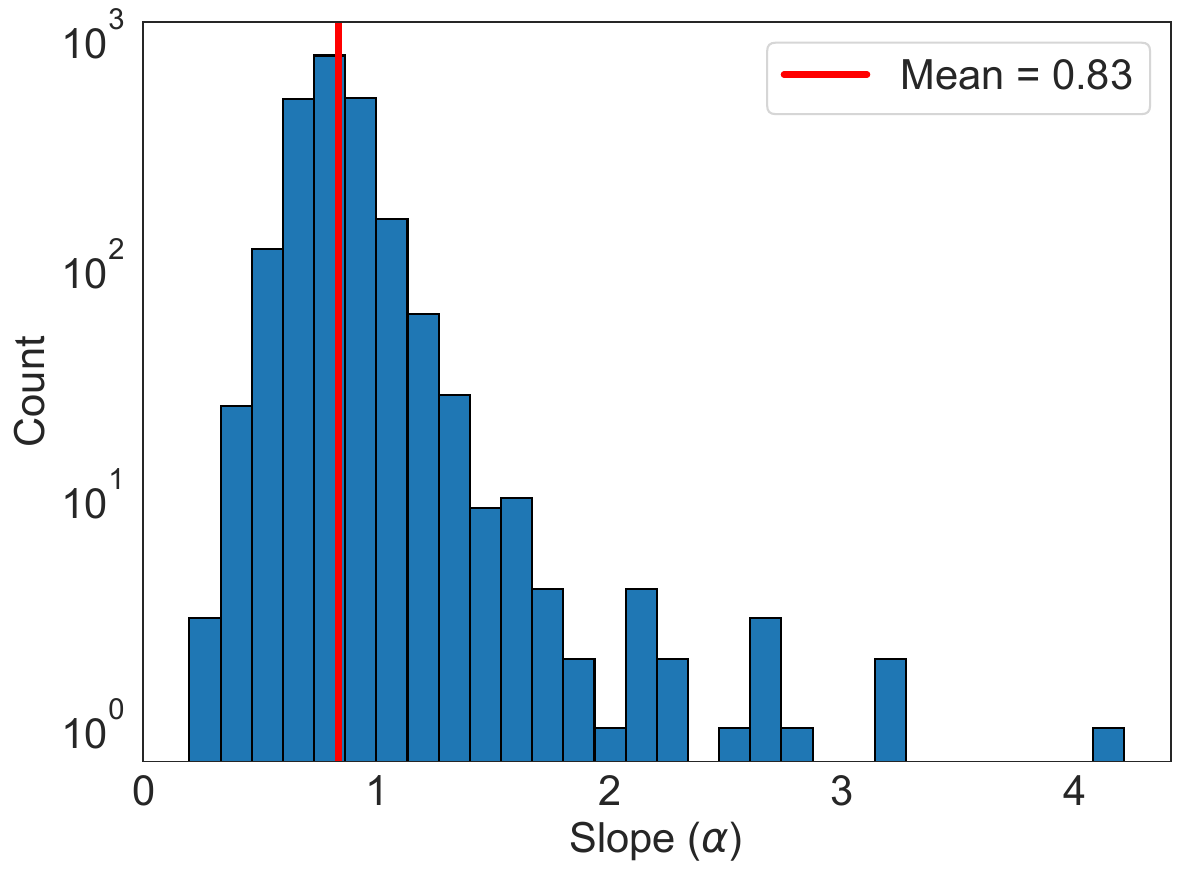}
    \caption{Distribution of all the slopes $\alpha$ obtained by performing a power-law fit on all the structures extracted recursively by decomposing each base level branch (from all available high-resolution column density maps) into its constituent children. The distribution peaks around 0.8, which shows a predominantly sub-linear $L-M$ relation for individual filamentary structures.}
    \label{fig:all_structures_histogram}
\end{figure}

Next, we repeat this analysis for all the branch structures identified in all high-resolution maps for all clouds, LOS and times. The full range of fitted slopes $\alpha$ is presented in the form of a histogram in Fig.~\ref{fig:all_structures_histogram}. We see a wide spread with a mean around 0.8.  This spread demonstrates that both the geometry of the filament and the spatial distribution of its mass (i.e., $N$ for the projected maps) and the resulting segmentation via the applied algorithms have a significant influence on the resulting $L-M$ relation. In consequence, the scaling observed for an individual filament is not universal, but instead reflects the complex interplay of hierarchical structure and physical conditions within the cloud.

\subsubsection{Explaining the \textit{L--M} relation for individual filaments}
\label{subsubsec:gravitational_fragmentation}

The hierarchical decomposition of individual filamentary structures (Figs.~\ref{fig:structures} and ~\ref{fig:all_structures_histogram})  demonstrates that inhomogeneities in the column density distribution lead to large variations in the $L-M$ scaling relation. To isolate the essential effects of inhomogeneity, we therefore turn to a simple toy model. This controlled framework allows us to mimic the segmentation of a filament and to directly evaluate the impact on the $L-M$ relationship. 

\begin{figure}
    \centering
    \includegraphics[width=1\linewidth]{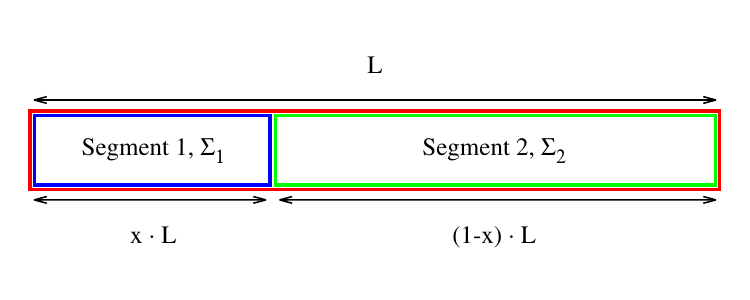}
    \caption{Toy model to demonstrate the origin of the $L-M$ scaling relation. The parent filament (red) is assumed to have a constant width everywhere and is segmented at a (variable) point located at a fraction $x$ of its entire length, $L$. Segments 1 (blue) and 2 (green) have uniform surface densities of $\Sigma_1$ and $\Sigma_2$, respectively. Subsequently, both segments are divided into a small number of sub-segments (see text).}
    \label{fig:toy_model}
\end{figure}

We begin by considering an idealized parent filament with a total mass $M_{\text{total}}$, the length, $L,$ and a constant width, $W$ (see Fig.~\ref{fig:toy_model}). This filament is segmented at a variable point on its longitudinal axis, producing two child filaments, segments 1 and 2 with the masses $M_1$ and $M_2$, respectively. We control the length of each segment in our toy model via a free parameter that we call `partition fraction' and denote it by $x$. The length of segment 1 is then $L_1 = x \cdot L$ and that of segment 2 is $L_2 = (1-x)\cdot L$.

We further assume that the mass distribution along the filament is not uniform, leading to different surface densities in each segment, $\Sigma_1$ and $\Sigma_2$, respectively\footnote{Within each segment we assume the surface density to be constant}. Using this setup, the masses of the two segments can be expressed as
\begin{align}
    M_1 &= \Sigma_1 \cdot x \cdot L \cdot W\\
    M_2 &= \Sigma_2 \cdot (1 - x) \cdot L \cdot W \, ,
\end{align}
with the total mass of the parent filament given by
\begin{align}
    M_{\text{total}} &= M_1 + M_2 \, .
\end{align}

The ratio of the surface densities of the two segments is defined as
\begin{equation}\label{eq:cd_ratio}
    R=\Sigma_1/\Sigma_2 \, .
\end{equation}
Without loss of generality, we can constrain ourselves to ratios $R \geq 1$.
Using this ratio $R$ and the partition fraction $x$, we can thus express the masses of the segments as
\begin{align}
    M_1 &= \frac{R \cdot x}{(1 - x) + R \cdot x} \cdot M_{\text{total}}, \\
    M_2 &= \frac{1 - x}{(1 - x) + R \cdot x} \cdot M_{\text{total}}.
\end{align}

To capture the hierarchical nature of filament fragmentation, each of the two segments is subsequently subdivided into a number of smaller substructures, randomly chosen between two and four. The lengths of these sub-segments are drawn from a Dirichlet distribution such that the sum of the subsegment lengths exactly reproduces the length of the parent segment. Their masses are then assigned assuming surface densities     scattered about that of the parent segment, with random variations of up to $\pm$30\% (allowing sub-segments to be somewhat denser or less dense than the parent), mimicking the realistic internal patchiness of filaments. The procedure is constructed in a way that it preserves both the total mass and length of segment 1 and 2, respectively, while introducing realistic inhomogeneities in the internal structure. All in all, the resulting toy filaments thus possess a hierarchical depth of two (parent $\rightarrow$ segments $\rightarrow$ sub-segments), thus matching the mean depth of 2.2 found for the base-level branches in the simulations (Sect.~\ref{subsec:hierarchical_decomp}).

To explore how segmentation impacts the inferred $L-M$ relation, we generate a population of such hierarchical filaments by systematically varying the partition fraction $x$ (drawn from a uniform distribution between zero and unity) and the surface density ratio $R$. For each realization of the fragmentation hierarchy we therefore obtain a set of structures consisting of the parent filament, its two primary segments, and the resulting sub-segments, each of them having its own length and mass. A power-law relation of the form $L \propto M^{\alpha}$ is then fitted to all structures in the hierarchy.

\begin{figure*}
    \centering
    \includegraphics[width=1\linewidth]{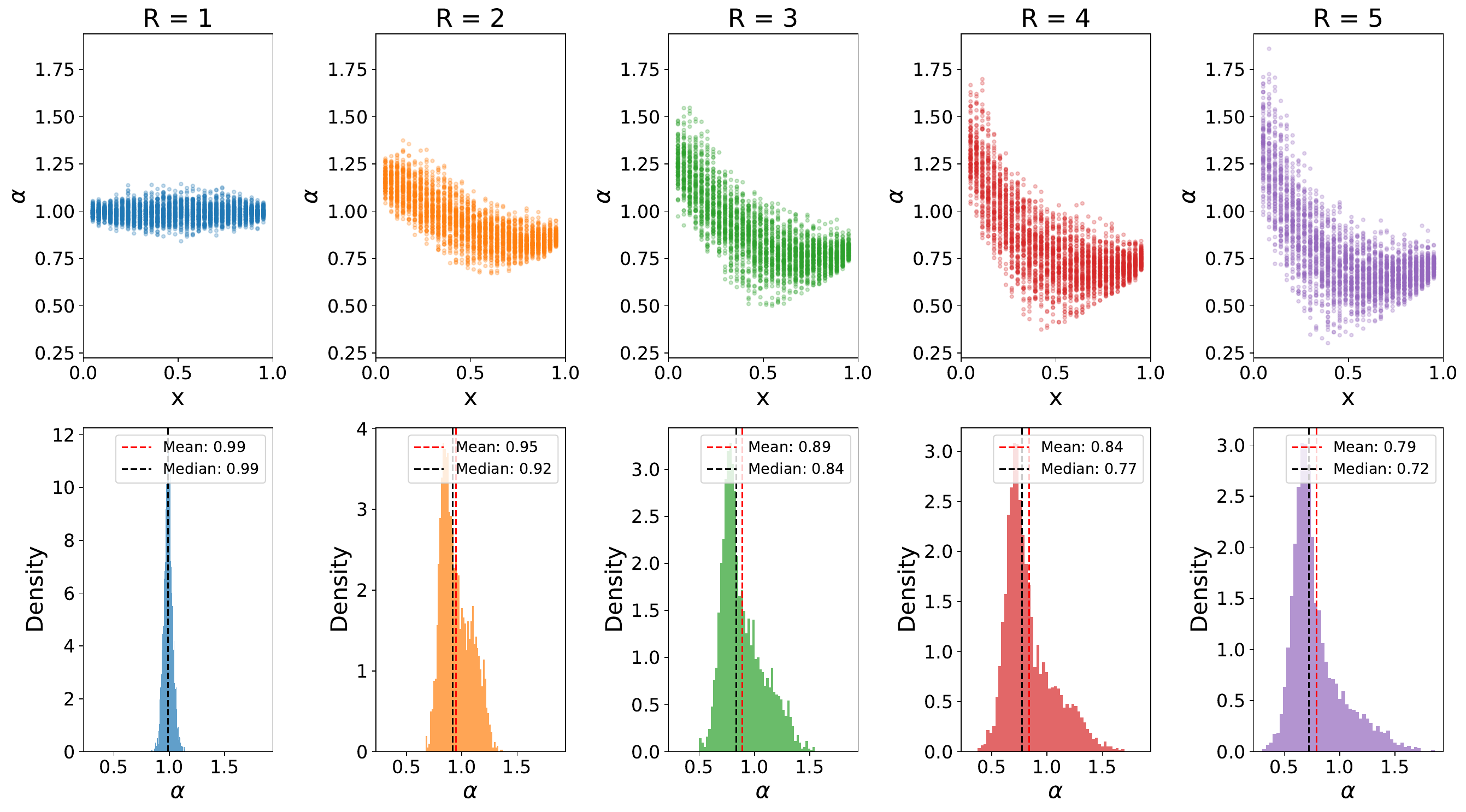}
    \caption{Power-law indices of the $L$-$M$ relation ($L \propto M^\alpha$) obtained from the filament fragmentation toy model (Fig.~\ref{fig:toy_model}). \textit{Top row}: Fitted slope $\alpha$ as a function of the partition fraction $x$ (the segmentation point along the parent filament) for various surface density ratios, $R$ (Eq.~\ref{eq:cd_ratio}).
\textit{Bottom row}: Corresponding PDFs of the fitted slopes. In the homogeneous case ($R=1$), the model recovers the trivial linear relation $L \propto M$ ($\alpha \approx 1$). As the surface density contrast increases ($R>1$), the distribution of slopes broadens and systematically shifts toward lower values. This demonstrates how increasing inhomogeneity in the mass distribution naturally produces a sub-linear $L$-$M$ scaling. In addition, the PDFs allow for a comparison with the actual distribution of $\alpha$ obtained from the simulations (Fig.~\ref{fig:all_structures_histogram}).
    }
    \label{fig:alpha_histogram_toy}
\end{figure*}

The resulting power-law index $\alpha$ is plotted as a function of $x$  for various values of $R$ in  the top row of Fig.~\ref{fig:alpha_histogram_toy}. Each point represents the slope obtained from fitting one hierarchical set of structures (parent, first-level segments, and their sub-segments).
When the mass distribution is uniform ($R=1$), the resulting relation approaches the trivial linear scaling $L \propto M$, corresponding to \mbox{$\alpha = 1$}\footnote{The scatter at a given value of $x$ is due to the imposed random fluctuations in the surface density for the sub-segments.}. However, when the mass distribution becomes inhomogeneous ($R \neq 1$), the inferred slope can deviate significantly from unity depending on how the filament is segmented (set by $x$) and how the mass is distributed along its length (set by $R$).

To build intuition for the distribution of $\alpha$ values obtained from all structures in all high-resolution column density maps (Fig.~\ref{fig:all_structures_histogram}), the bottom row of Fig.~\ref{fig:alpha_histogram_toy} shows the PDFs of the fitted slopes derived from the toy model. For each $R$, the PDF consistently peaks at $\alpha < 1$, and this peak systematically shifts toward lower $\alpha$ values with increasing $R$ -- that is, as the column density becomes more inhomogeneous. Specifically, the mean slope decreases from $\alpha = 0.95$ for $R=2$ to $\alpha = 0.79$ for $R=5$, illustrating how internal mass inhomogeneity drives a sub-linear $L-M$ scaling. At the same time, the spread in $\alpha$ also increases as $R$ increases. In other words, greater inhomogeneity results in stronger deviations from linearity ($\alpha = 1$).

Finally, trying to match the distribution of $\alpha$ of our toy model to the actual distribution derived from the simulations (Fig.~\ref{fig:all_structures_histogram})
would imply surface density ratios of the order of $R \simeq 3-5$. This exceeds the contrast actually measured between the substructures extracted from the simulations: for the direct children of a branch (i.e., a split), we first calculate the mean surface density of each child, $\langle \Sigma \rangle_{\textrm{child}}$. Next, taking the ratio between the maximum and minimum $\langle \Sigma \rangle_{\textrm{child}}$, i.e., $R$, for each split yields a mean (median) of $\simeq$1.5 ($\simeq$1.2), and only $\sim$3\% of all splits reach $R = 3-5$; this holds equally for the first split of a base-level filament and for deeper splits. The discrepancy reflects the idealizations of the toy model: its filaments are straight, of uniform width and uniform surface density within each segment, and their second hierarchical level is generated by a simple, mass- and length-conserving partition. In addition, real filamentary structures can be curved (in both the plane of sky as well as along the LOS) and their substructures are not necessarily aligned with their parent, and their measured masses include fore- and background material along the line of sight; all that affects the measured $L-M$ scaling and is absent from the model. The toy model should therefore be read as demonstrating the mechanism that inhomogeneity in the surface density drives $\alpha$ below unity, rather than as a prediction of the contrast $R$ found in filamentary substructures.

We also note here two caveats of our toy model. First, $x$ is uniformly distributed in our model. This general assumption yields a broad range of mean slopes. In contrast, drawing $x$ from a non-uniform distribution centered around 0.5 (e.g., a normal distribution corresponding to roughly equal-length segments) would shift the resulting PDFs for each $R$ toward slightly lower mean values around $\alpha \approx 0.5$. This can also be seen by the fact that $\alpha$ typically has its lowest  value around $x \simeq 0.5$ (top row of Fig.~\ref{fig:alpha_histogram_toy}).  Second, it should also be noted that we have assumed a constant width for the segments. However, varying the widths instead of $R$ would produce qualitatively similar results, i.e., a wide range of $\alpha$.

To summarize, this toy model illustrates how observed deviations from a linear $L-M$ scaling in highly resolved, individual filamentary structures can naturally arise due to (i) the actual point of segmentation and (ii) a non-uniform mass distributions (i.e., varying column density). It thus provides an interpretive framework for the diverse slopes obtained in our hierarchical filament analysis (Figs.~\ref{fig:structures} and~\ref{fig:all_structures_histogram}).

\section{Discussion}\label{sec:discussion}

\subsection{Effectiveness of the filament identification process}

The RHT effectively encapsulates almost all the filamentary features in the considered MCs with the given set of input parameters, as shown in Fig.~\ref{fig:rht_and_dendrogram} and \ref{fig:selected_structure}. This is supported by our finding that filamentary structures account for approximately 20-30\% of the total gas mass in each cloud (Sect.~\ref{subsec:data_prep}), and by the presence of filaments spanning over two orders of magnitude in column density. This aligns well with the findings of \citet{arzoumanian_2019}, where the authors found that more than 15\% of the total gas mass in the clouds is found to be in the form of filamentary structures.

We chose to apply a dendrogram analysis to the RHT output rather than to the corresponding column density maps, which departs from the conventional approach, where filament identification algorithms are directly applied to column density/intensity maps \citep[e.g.,][]{arzoumanian_2011, schisano_2014, koch_2015, arzoumanian_2019, kumar_2020, carriere_2022, kirk_2024}. The key advantage of this strategy is the ability to identify filaments across all spatial scales, irrespective of their column densities. The RHT is inherently sensitive to elongated morphology and performs equally well in both high- and low-column density regions. This advantage is clearly reflected in the broad range of filament column densities seen in Fig.~\ref{fig:M-L_trunk_1024_res_all_structures}.
Without the aid of RHT, a direct dendrogram analysis on the column density map would primarily respond to column density peaks rather than linear coherence, causing faint or diffuse filaments to be merged into larger structures or lost within hierarchical branches, thereby under-representing the true filamentary population.

\subsection{Hierarchical fragmentation modes} \label{subsec:filaments_slopes_at_ind_level}

Our hierarchical decomposition performed on the highest-resolution maps (Fig.~\ref{fig:structures}) shows that a single branch filament can break into leaves and sub-branches with different $L-M$ scaling relations. This is exactly what one expects if the mass distribution and geometry of the parent is not uniform due to the combination of turbulence, self-gravity, magnetic field, as well as  ongoing accretion onto the filament (and potential dispersal).

In the literature, two approaches are discussed which could explain this observed hierarchical structure. 
First, large, parsec-scale parent filaments host sub-parsec filaments which, in turn, contain fiber-like substructures. These assemble and disperse on dynamical timescales and fragment unevenly as turbulence seeds multi-scale perturbations, which are subsequently amplified by gravity  \citep{Padoan_2002, Hennebelle_2012, Kraljic_2014}. This top-down fragmentation picture (``filaments $\rightarrow$ fibers $\rightarrow$ cores'') is well established in Taurus and other nearby regions \citep{Hacar_2013, hacar_2018}. 

In contrast to that, the bottom-up model by \citet{smith_2014} shows that filamentary structures in turbulent molecular clouds can also arise from a network of small, coherent sub-filaments produced by turbulent stagnation points. These sub-filaments typically have sub-parsec scales, implying that larger-scale filaments are created by merging of these small-scale filaments during gravitational collapse. The larger (parsec-scale) filaments thus correspond to bundles of smaller fibers whose hierarchical and time-variable structure reflects the turbulent dynamics of the cloud. 

To differentiate these two scenarios, filamentary structures need to be tracked over time, which is not within the scope of this work. We note that \citet{Feng_2024} follow the evolution of individual filaments over time. The authors find different relations in the $L-M$ plane as time evolves (corresponding to a wide range of $\alpha$ in the context of our work), similar to our variations in the $L-M$ relation at fixed time but varying scales (see Fig.~\ref{fig:structures}). This striking similarity indicates that strong variations in the scaling relation at the level of individual structures occur both in time and over scales.

\subsection{Turbulent fragmentation: A random walk} \label{subsubsec:turbulent_fragmentaition}
In Sect.~\ref{subsubsec:consistency_across_ind_fils}, a sub-linear trend in the $L-M$ relation was found when an individual filamentary structure (identified at a coarse resolution of 0.96~pc) is tracked across resolutions down to 0.12~pc (see Figs.~\ref{fig:selected_structure} and~\ref{fig:selected_structure_M_L_plot}). The inferred $L-M$ scaling with $\alpha$ close to 0.5 is remarkably similar to that inferred for individual clouds observed at different resolutions \citep[see Fig.~9 in][]{hacar_2023}.  Our results thus confirm these observational results from a simulation perspective.

As suggested by \citet{hacar_2023}, these findings can be explained if filament formation follows a random-walk process across scales: a large-scale, self-gravitating, initially linear filament of total length $L_0$ and mass $M_0$, embedded in a turbulent MC, is subject to bending and internal density perturbations. The latter are amplified by self-gravity, ultimately leading to hierarchical fragmentation. Following \citet{hacar_2023}, we assume that the parent filament fragments into $N$ sub-filaments due to turbulence. Unlike for pure segmentation (the toy model case, Sect.~\ref{subsubsec:gravitational_fragmentation}), the total length of these sub-filaments $i$ can exceed the length of the main filament, i.e., $\sum_i L_i > L_0$. In a random-walk picture, each sub-filament acquires a characteristic length of $L=L_0/\sqrt{N}$ and a mass of $M=M_0/N$, respectively. Eliminating $N$ thus yields the relation
\begin{equation}
    L = L_0 \left(\frac{M}{M_0}\right)^{1/2} \quad \Rightarrow \quad L \propto M^{1/2},
\end{equation}
which provides a straightforward argument for the emergence of a sub-linear slope in the $L-M$ relation.

We emphasize that this is a geometric, statistical argument. In the toy model proposed by \mbox{\citet{hacar_2023}}, turbulent bending of a parent filament is treated as a random-walk process, implying that its $N$ sub-filaments each span $L_0/\sqrt{N}$, so the mass is divided faster ($M_0/N$) than the length. The authors further note that the same argument can be made for a bottom-up (`fray and gather') scenario. We therefore do not claim that the substructures in our simulations are produced specifically by turbulent fragmentation: distinguishing between top-down and bottom-up origins requires following structures in time, which is beyond the scope of this paper (see Sect.~\ref{subsec:filaments_slopes_at_ind_level}).

This interpretation is supported by our results when considering individual filamentary structures across various resolutions (see Figs.~\ref{fig:selected_structure} and~\ref{fig:selected_structure_M_L_plot}), where we see the sub-linear scaling with $\alpha$ close to 0.5. In addition, the smaller-scale filaments seen in Fig.~\ref{fig:selected_structure} partly deviate from the direction of the overarching structure, similar to what the random walk model predicts. Hence, the random-walk fragmentation scenario provides an explanation, wherein the effective scaling emerges from the interplay between turbulence and gravity in redistributing mass along a bending, fragmenting structure.

\subsection{The sub-linear \texorpdfstring{$L-M$}{L-M} relation: An inherent feature of the ISM} \label{subsec:ensemble_sub_lin}

In Sect.~\ref{subsubsec:consistency_across_ind_fils} we showed that, when the resolution is increased, finer substructures appear within a single filament. The resulting $L-M$ relations have power-law slopes of \mbox{$\alpha = 0.45 - 0.70$} (Fig.~\ref{fig:selected_structure_M_L_plot}). In addition, when we consider the ensemble of all these filamentary structures (at all resolutions) in a single $L-M$ plot, we also obtain a slope of $\alpha \simeq 0.5$ (see Fig.~\ref{fig:M-L_baselevel_all_res}). Similarly, when considering filaments at the highest resolution only, again a sub-linear relation with $\alpha \simeq 0.5$ is found (see Fig.~\ref{fig:M-L_trunk_1024_res_all_structures}) -- despite the range of scalings for individual filaments (see Fig.~\ref{fig:structures}). Taken together, these results indicate that the $\alpha \approx 0.5$ scaling is an intrinsic outcome of hierarchical fragmentation, rather than an effect of resolution or other observational constraints.

As discussed in \citet{hacar_2023}, this scaling can also be understood in the context of the third Larson relation $n \propto L^{-1.1}$ \citep{larson_1981}, which implies $M \propto L^{1.9}$ and therefore $L \propto M^{0.53}$. The resulting relation is close to the $L\propto M^{0.5}$ scaling measured here and reflects the hierarchical nature of molecular clouds. However, Larson's relation implies almost constant column densities. As shown in Fig.~\ref{fig:M-L_trunk_1024_res_all_structures}, we see substantial variations in $\langle N \rangle$ over approximately two orders of magnitude, which implies that the situation is more complex. Further support for this is given by our results of the toy model, showing that inhomogeneity in the surface density is one of the main drivers of the sublinear $L-M$ relation (see Fig.~\ref{fig:alpha_histogram_toy}).

Last but not least, we compare our results in more detail with those of \citet{Feng_2024}, whose work is close in scope to ours. The authors extract simulated MCs from the galactic-scale `Cloud Factory' simulations \citep{Smith_2020}, and identify filaments from synthetic dust continuum maps with the \texttt{FilFinder} algorithm -- a methodology which differs from ours in the underlying simulations, the map generation, and the identification technique. Despite these differences, they recover a sub-linear ensemble relation of $L \propto M^{0.45}$, in very good agreement with our mean value of $\langle \alpha \rangle \simeq 0.53$ (Sect.~\ref{subsec:hierarchical_decomp}) and with the observational value of $\alpha \approx 0.5$ \citep{hacar_2023}. The sub-linear $L-M$ relation appears thus to be robust against the choice of simulation, identification algorithm, and analysis strategy.

Moreover, the two studies probe complementary aspects of the problem: \citet{Feng_2024} follow individual filaments over time and attribute their motion in the $L-M$ plane to accretion, segmentation, and dispersal (see their Fig.~10), whereas we decompose the filament population at fixed evolutionary stages into its hierarchical constituents, both across resolutions and within the highest-resolution maps. In particular, our fixed-time hierarchical decomposition can be read as an instantaneous counterpart of the segmentation channel identified by \citet{Feng_2024}. The fact that both the temporal evolution of individual filaments \citep{Feng_2024} and the instantaneous hierarchical decomposition (this work) produce a large diversity of individual scaling behaviors around a robust, sub-linear ensemble average (see also Sect.~\ref{subsec:filaments_slopes_at_ind_level}) strongly supports the interpretation that the $L \propto M^{\sim0.5}$ relation is an inherent, statistical property of the hierarchically structured, filamentary ISM, rather than the signature of one specific evolutionary path. Finally, \citet{Feng_2024} report that projection effects systematically affect the measured masses and lengths of individual filaments, particularly in crowded regions, fully in line with the discussion on individual structures in Sect.~\ref{subsubsec:hierarchy_in_ind_fils}. 

\section{Conclusion}\label{sec:conclusion}

In this work, we present a systematic study of filament identification and characterization in high-resolution MC simulations within the SILCC-Zoom project \citep{seifried_2017}. We introduce a two-step identification method that first uses the Rolling-Hough Transform (RHT) to highlight coherent linear structures. Second, we apply a dendrogram-based segmentation to the RHT output to extract individual filamentary structures as hierarchical branches and leaves. We apply this method to column density maps of multiple MCs at different evolutionary stages and across four spatial resolutions (0.12, 0.24, 0.48 and 0.96~pc).

For each identified filamentary structure, we extract basic quantities such as its length, $L$, mass, $M,$ and average column density and analyze their distribution.  We focus mainly on the $L-M$ relation as compiled in \citet{hacar_2023}. The main results of our analysis are summarized in the following:

\begin{enumerate}
    \item We show that the combined approach of the RHT and dendrogram analysis reliably recovers filamentary structures with a variety of morphologies. The average column densities span a wide range with values extending over approximately" two orders of magnitude from $\sim$10$^{21}$ to $\sim$10$^{23}$ cm$^{-2}$. Our method is thus particularly effective in revealing faint, low-contrast filaments that other methods often miss, which enables a more complete and physically meaningful census of filamentary structures in the simulated MCs. 

    \item The $L-M$ relation generally follows a sub-linear power law $L \propto M^\alpha$ across the ensemble of hierarchical filamentary structures found in this work.
    Including filamentary structures across the various resolutions ($0.12 - 0.96$~pc), we find typical power-law indices for different clouds of $\langle \alpha \rangle = 0.54 \pm 0.07$ (Fig.~\ref{fig:M-L_baselevel_all_res})  in good agreement with \citet{hacar_2023}.

    \item Considering only filamentary branch and leaf structures at the highest spatial resolution (0.12~pc) yields a very similar mean slope of 
    $\langle \alpha \rangle  = 0.53 \pm 0.07$ (Fig.~\ref{fig:M-L_trunk_1024_res_all_structures}). 
    The consistency of these values demonstrates that the sub-linear $L-M$ scaling is not a by-product of resolution blending, but an intrinsic statistical property of the turbulent, hierarchically structured ISM.

    \item When tracking individual filaments across different resolutions (Fig.~\ref{fig:selected_structure_M_L_plot}),
    the resulting power-law indices range from $\alpha \simeq 0.45$ to $0.70$. This is consistent with the     geometric expectation $L \propto M^{1/2}$ of a fragmentation-driven random walk model \citep{hacar_2023}. We note that this is offered as an interpretation for the slope across resolutions, not as evidence of a specific (turbulent versus gravitational) fragmentation mechanism.

    \item In contrast to the robust ensemble average stated before, the $L-M$ relations of individual branch filaments identified and decomposed at the highest resolution exhibit substantial diversity with $\alpha$ ranging from $\sim0.2$ to $\sim3$ (Fig.~\ref{fig:all_structures_histogram}).
    We find examples of super-linear $(\alpha > 1)$, near-linear $(\alpha \simeq 1)$, and sub-linear $(\alpha < 1)$ $L-M$ relations, which indicates that the $L-M$ scaling can vary strongly at the local level. 
    
    \item We explain this variability with a simple toy model, which demonstrates that the slope $\alpha$ is highly sensitive to both the filament’s internal mass distribution (degree of inhomogeneity) and the geometry of its segmentation. 
    Comparing mean surface densities of filamentary substructures in our simulation yields modest variations of 50\%.

    \item 
    Our toy model of filament segmentation also shows that the distribution of power-law indices peaks at $\alpha < 1$ (Fig.~\ref{fig:alpha_histogram_toy}), with the peak shifting as the inhomogeneities in the filaments increase. This behavior provides further quantitative support to our interpretation that the sub-linear $L-M$ relation, i.e., $L \propto M^{\sim0.5}$, found for the ensemble of filamentary structures reflects a statistical consequence of hierarchical fragmentation.
\end{enumerate}

Taken together, the various results, which all show an average sub-linear power-law relation $L \propto M^{\alpha}$ with $\alpha \simeq 0.5$, indicate that the emerging $L-M$ relation, as compiled in \citet{hacar_2023}, is a robust property of the hierarchical structure and the fragmentation process that occur in MCs. In future work we plan to investigate further properties of the filamentary structures such as the width and velocity dispersion. We also intend to perform a stringent characterization of the wide range of morphological shapes of structures found in this work.

\section*{Data availability}

The data underlying this article can be shared for selected scientific purposes after request to the corresponding author.

\begin{acknowledgements}
RP acknowledges support from the Deutsche Forschungsgemeinschaft (DFG, German Research Foundation) under Grant No. BA3706/19-1 and Germany’s Excellence Strategy (EXC 2121 “Quantum Universe”, 390833306).
DS acknowledges support of the Bonn-Cologne Graduate School, which is funded through the German Excellence Initiative as well as funding by the DFG via the Collaborative Research Center SFB 1601 ``Habitats of massive stars across cosmic time'' (subprojects B1 and B4).
Furthermore, the project has received funding from the program “Profilbildung 2020", an initiative of the Ministry of Culture and Science of the State of Northrhine Westphalia.
AH acknowledges support from the European Research Council (ERC) under the European Union’s Horizon 2020 research and innovation program (Grant agreement No. 851435).
\end{acknowledgements}

\bibliographystyle{aa}
\bibliography{Bibliography_filament}

\begin{appendix}

\section{Parameter optimization}\label{app:para_opt}

Here, we describe the methodologies employed to determine the optimal set of parameters for both the RHT and \texttt{AstroDendro} routine using the highest resolution column density maps. The selection was based on systematic variation, visual inspection, and evaluation of the output quality across different configurations, ensuring robust detection of filamentary structures while minimizing noise and artifacts.

\subsection{Optimization of RHT parameters} \label{subapp:opt_rht}
We describe the procedure followed to determine the optimal set of parameters for the RHT. As outlined in Sect.~\ref{subsec:RHT}, the RHT algorithm requires three input parameters: \textit{wlen}, \textit{smr}, and \textit{frac}. We begin with the default values (\textit{wlen} = 55, \textit{smr} = 15, and \textit{frac} = 0.7) and apply the RHT to the column density map projected along the $z$-direction for MC1-HD at $t_{\mathrm{evol}} = 2.5~\mathrm{Myr}$. Upon overlaying the RHT results on the column density map (not shown), it becomes evident that the initial parameter choice is not well suited for capturing the smaller-scale filamentary structure present in the map. In particular, the large window length and strong smoothing suppress fine structure and lead to an overly coarse representation of the filament network.

\begin{figure}
    \centering
    \includegraphics[width=1\linewidth]{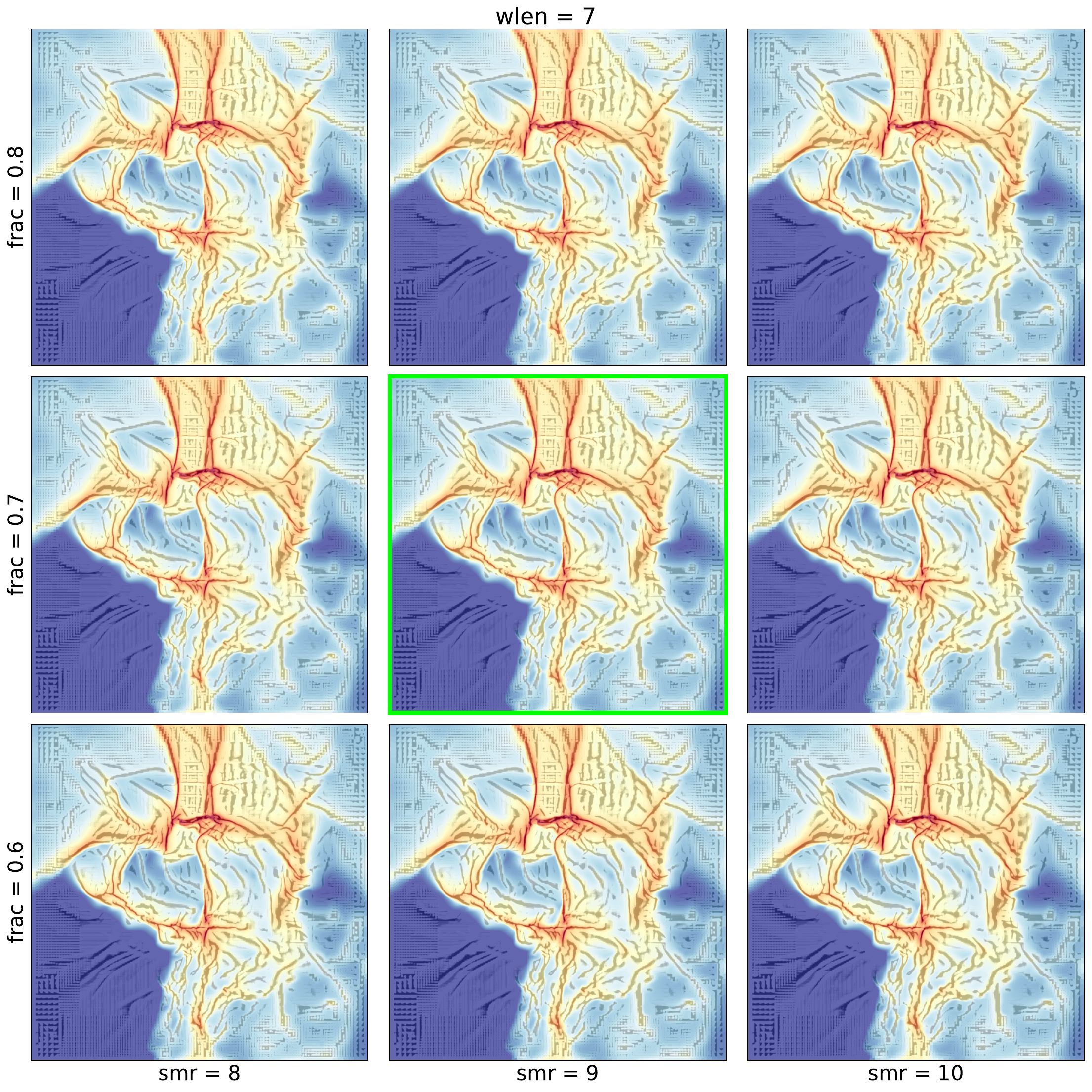}
    \caption{RHT outputs for different combinations of \textit{smr} and \textit{frac}, with fixed \textit{wlen} = 7, overlaid on the column density map of run MC1-HD at $t_{\mathrm{evol}} = 2.5~\mathrm{Myr}$ projected along the $z$-direction. We chose the parameter set \textit{wlen} = 7, \textit{smr} = 9, and \textit{frac} = 0.7 for our work as it lies within the stable region of
    parameter space.}
    \label{fig:wlen_7}
\end{figure}

To address these issues, we systematically vary the parameters and find that for the following values 
\begin{itemize}
    \item $\mathit{wlen} \in [3, 5, 7, 9]$
    \item $\mathit{smr} \in [8, 9, 10]$
    \item $\mathit{frac} \in [0.6, 0.7, 0.8]$
\end{itemize}
the identified filamentary structures are remarkably robust and only very minor differences arise between individual combinations. These variations are subtle and, in most cases, not noticeable to the naked eye, as can be verified from Fig.~\ref{fig:wlen_7}. This indicates that the RHT solution is in a regime in which the global filamentary topology and connectivity remain stable under moderate parameter variations.
Based on this analysis, we adopt the parameter set \textit{wlen} = 7, \textit{smr} = 9, and \textit{frac} = 0.7. This choice lies within the stable region of parameter space and balances the need to detect fine structures while avoiding the inclusion of artifacts. Applying this combination to other column density maps consistently results in reliable and reproducible filamentary structures.

\subsection{Optimization of \texttt{AstroDendro} parameters}\label{subapp:opt_astrodendro}

\begin{figure}
    \centering
    \includegraphics[width=\linewidth]{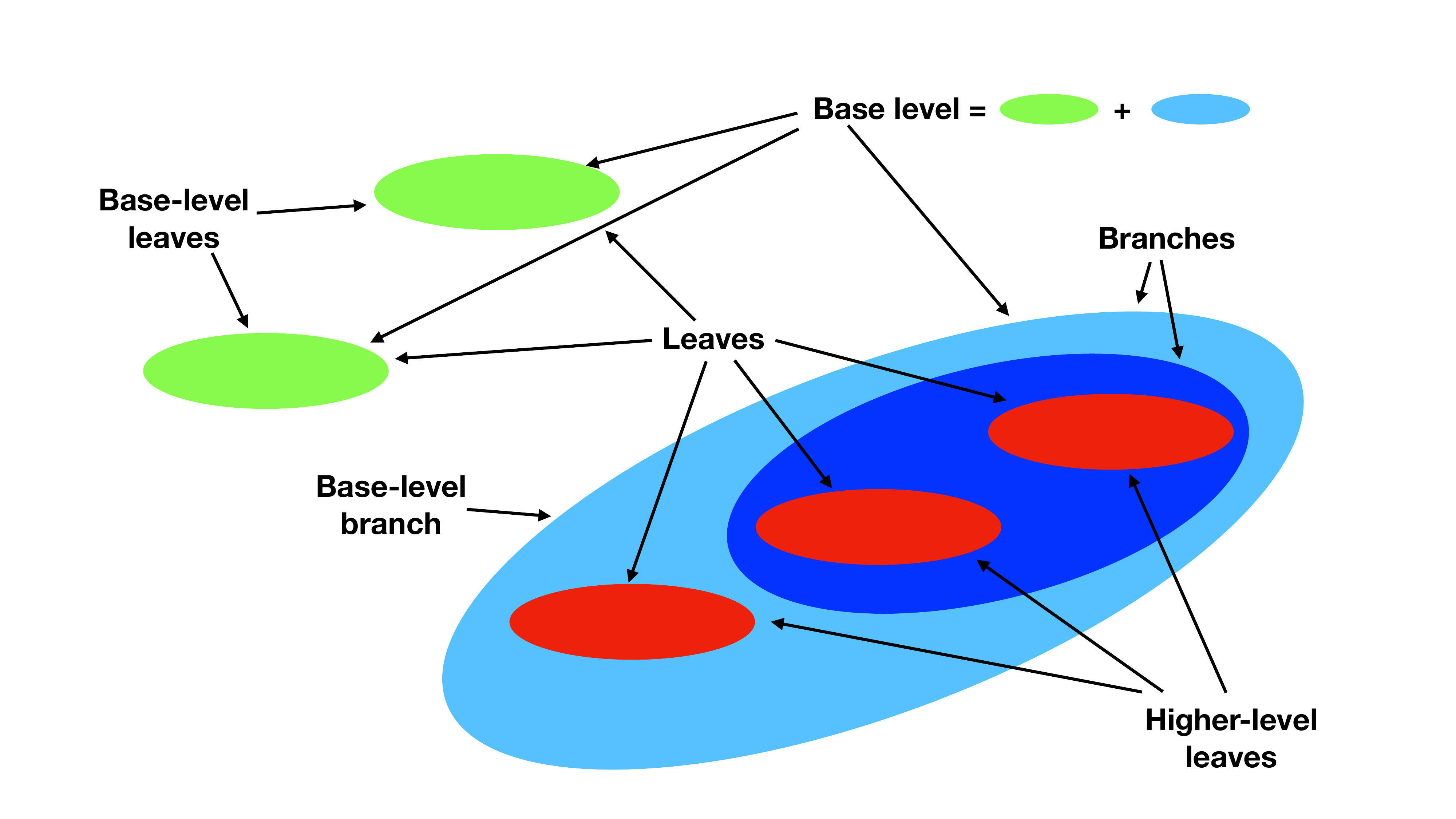}
    \caption{Definition of the terminology used in this work for the hierarchical structures identified by the dendrogram algorithm.}
    \label{fig:definition}
\end{figure}

The methodology adopted to identify the optimal set of parameters for \texttt{AstroDendro} applied to the RHT maps (see Sect.~\ref{subsec:dendrogram}) is described in the following. Before doing so, we briefly clarify the terminology that we use in this work to denote the different parts of the hierarchical structure determined by the dendrogram routine. This is done graphically in Fig.~\ref{fig:definition}. Beside the usually used terms for branches, which contain sub-branches, and leaves, which do not contain substructures, we particularly refer to the base level, which comprises the base-level leaves and branches, i.e., those structures identified at the lowest iso-contour given by the value of $min\_value$  (see below). Base-level branches correspond to what is frequently called `trunks' in the literature \citep[e.g.,][]{rosolowsky_2008}.

This procedure is carried out for the highest-resolution maps only; for lower resolutions, some parameter values were adapted accordingly (see Appendix~\ref{app:choice_of_min_val}). We begin with general considerations before systematically exploring the influence of individual parameters. 

\begin{figure}
    \centering
    \includegraphics[width=0.50\linewidth]{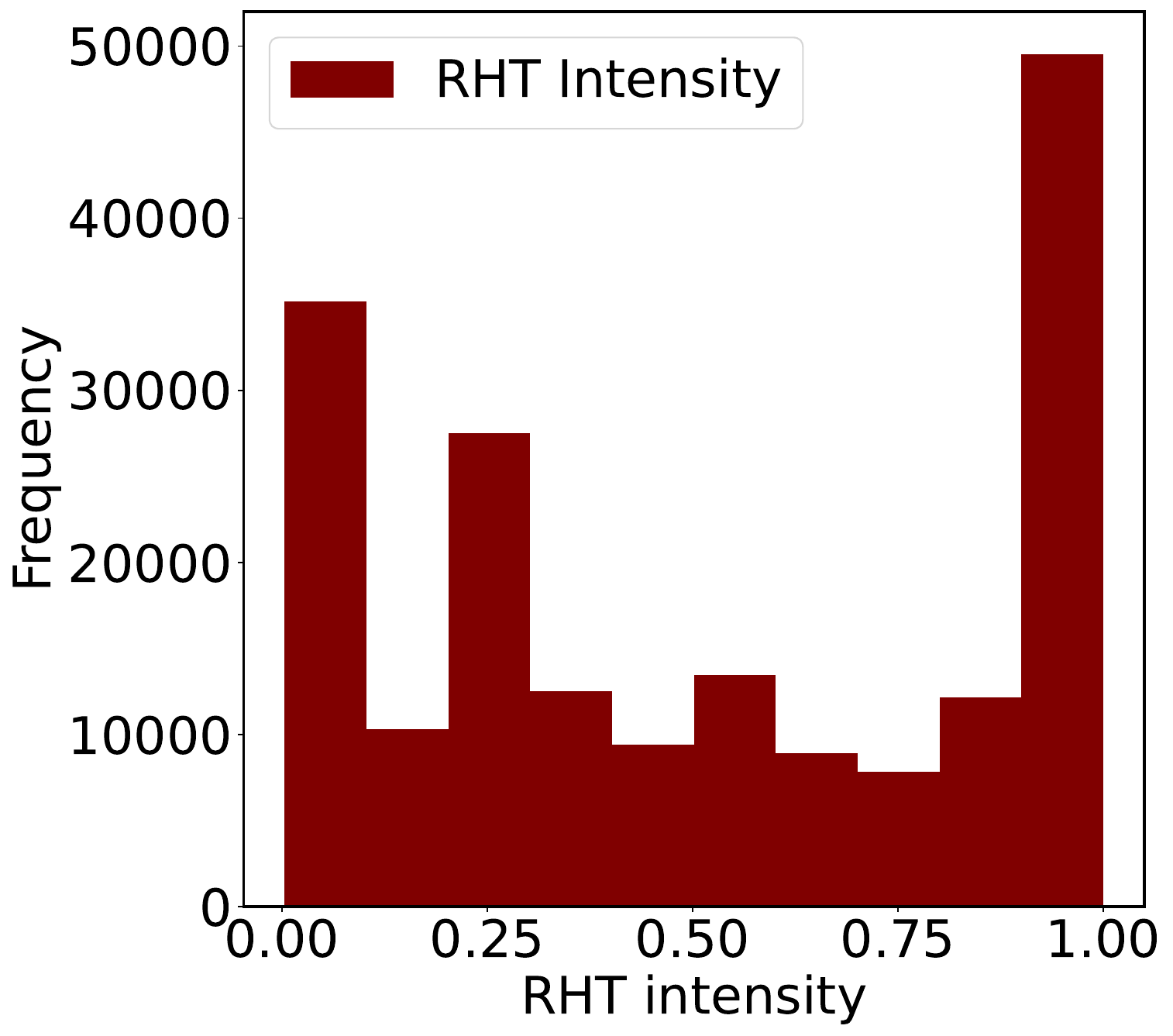} 
    \includegraphics[width=0.46\linewidth]{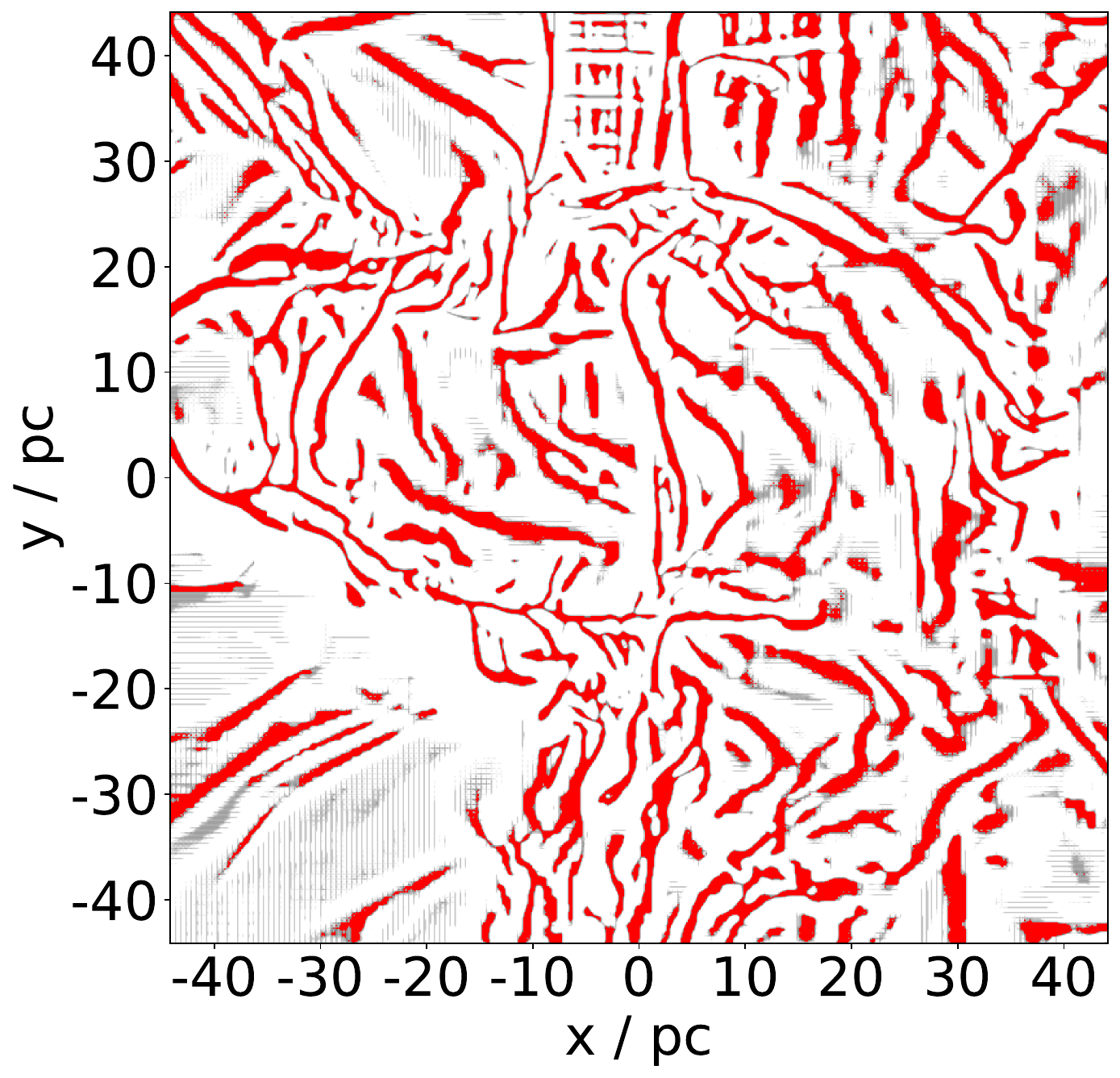} 
    \caption{\textit{Left}: Histogram of RHT values obtained from the RHT analysis of the column density map of the MC1-HD simulation at $t_{evol}=2.5~\mathrm{Myr}$ along the $z$-direction. The prominent peak near 0.9 indicates the presence of well-defined filamentary structures. \textit{Right}:  RHT output value below a threshold of 0.5 (pixels above this value are shown in red). These low-value structures primarily correspond to artifacts or peripheral pixels near filament boundaries, and are not associated with coherent filamentary structures.}
    \label{fig:RHT_hist_threshold}
\end{figure}

To determine the optimal value for the \textit{min\_value} parameter, we first analyze the RHT output. Since the RHT output can be interpreted as a measure of likelihood of a structure being identified as a filament, genuine filamentary structures are expected to exhibit higher values, while grid artifacts tend to have somewhat lower values. A histogram of the RHT values (see left panel of Fig.~\ref{fig:RHT_hist_threshold})
shows a clear peak at values above 0.8. We interpret pixels with such high values to be located within filaments, whereas pixel with lower values most likely belong to a mixture of filamentary features and artifacts.

To evaluate the trade-off between completeness and contamination, we showed the RHT output in Fig.~\ref{fig:RHT_hist_threshold}, but discard all pixels which have values above 0.5 (shown in red).
As can be seen, most of the artifacts visible in the RHT map have values below 0.5, whereas the main filamentary structures are mostly found at values above 0.5.
Hence, in the following we used values of \textit{min\_value} above this threshold of 0.5.
Furthermore, although placing the threshold near the histogram peak at 0.9 might eliminate almost all artifacts, it also risks excluding many small or faint filaments that contribute to the structural complexity of the medium.
Therefore, we varied \textit{min\_value} within the range [0.5, 0.9] to capture this trade-off systematically.

Next, we consider the \textit{min\_delta} parameter, which sets the minimum contrast required to distinguish nested substructures. While \textit{min\_delta} can technically range up to 1 in our case, we found that values above 0.4 significantly suppress the identification of independent substructures across our chosen \textit{min\_value} range. High \textit{min\_delta} values tend to merge nearby peaks into a single parent structure, thereby reducing the total number of detections and thus oversimplifying the complex filamentary structure of the ISM.

We fix the \textit{min\_npix} parameter -- the minimum number of contiguous pixels required for a structure to be considered valid -- to 10. This choice filters out isolated noise peaks but retains small-scale filamentary structures that are scientifically relevant to this study.

With these pre-considerations in mind, we systematically (in steps of 0.1) vary \textit{min\_value} in the range  [0.5, 0.9] and \textit{min\_delta} in the range [0, 0.4]. The left panel Fig.~\ref{fig:heatmaps} summarizes the parameter-space behavior of the total number of base-level structures detected. It decreases with increasing \textit{min\_delta}, as more and more substructures get merged into their parents. Conversely, the number of structures increases with higher \textit{min\_value}, as more diffuse bridges between peaks are suppressed, leading to a greater number of isolated detections . We note that, by construction, \textit{min\_delta} cannot exceed $1-\textit{min\_value}$; the parameter-space exploration is restricted accordingly, which is why the upper-triangular cells in Fig.~\ref{fig:heatmaps} are not populated.

\begin{figure*}
    \centering
    \includegraphics[width=1\linewidth]{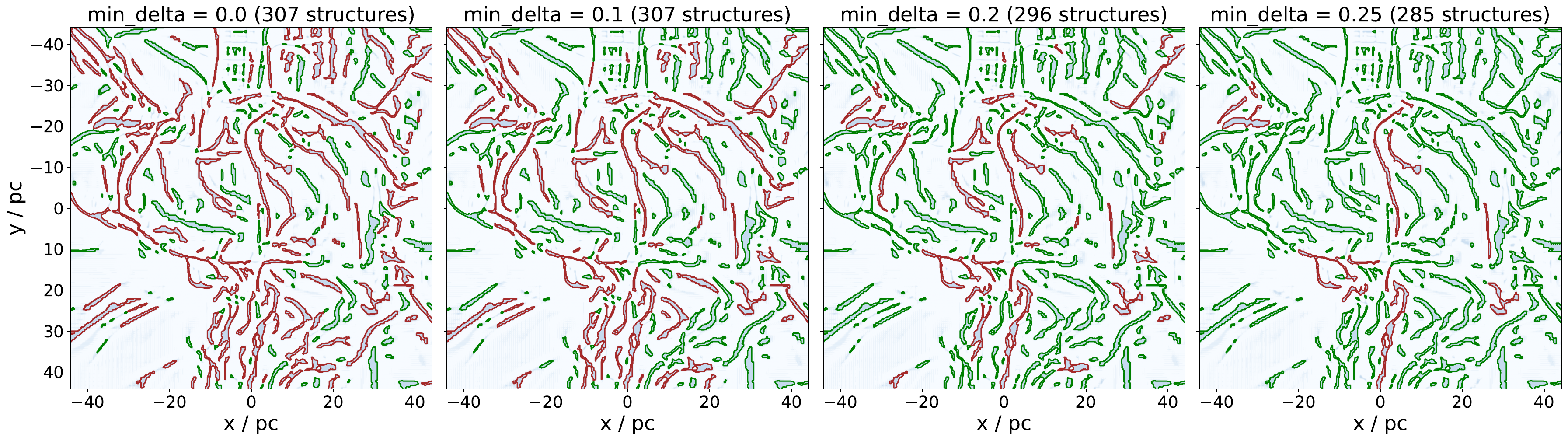}
    \caption{Contours of the identified base-level structures (leaves in green, branches in brown) for increasing values of \textit{min\_delta}  (left to right) at fixed \textit{min\_value} = 0.7 and \textit{min\_npix} = 10, applied to the RHT output of MC1-HD at $t_\mathrm{evol}$ = 2.5~Myr projected along the $z$-direction. With increasing \textit{min\_delta}, substructures are progressively merged into their parent structures (branches turn into leaves), erasing the hierarchical information, while the total number of structures decreases only mildly.}
    \label{fig:min_delta_var}
\end{figure*}

\begin{figure}
    \centering
    \includegraphics[width=0.49\linewidth]{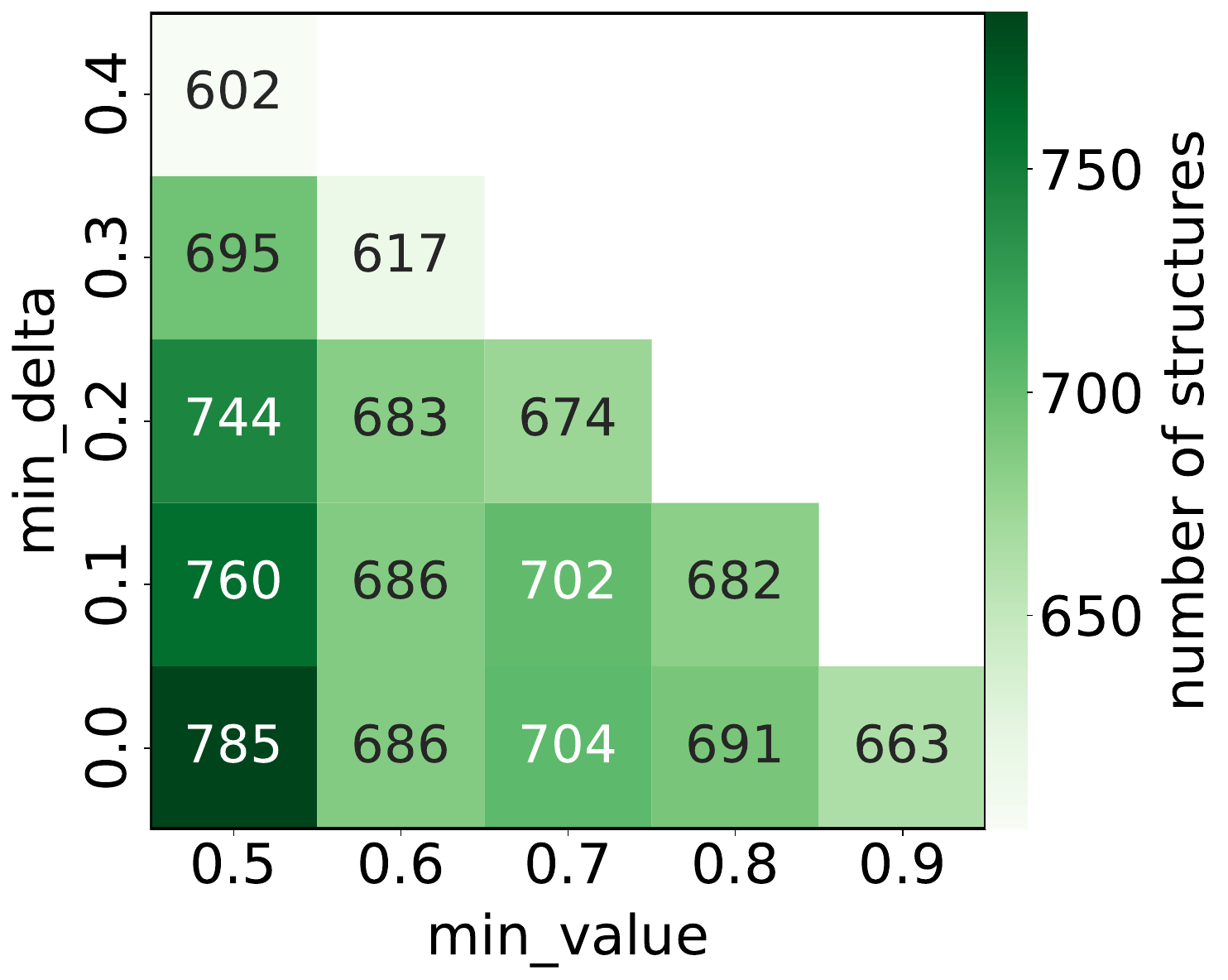} 
    \includegraphics[width=0.49\linewidth]{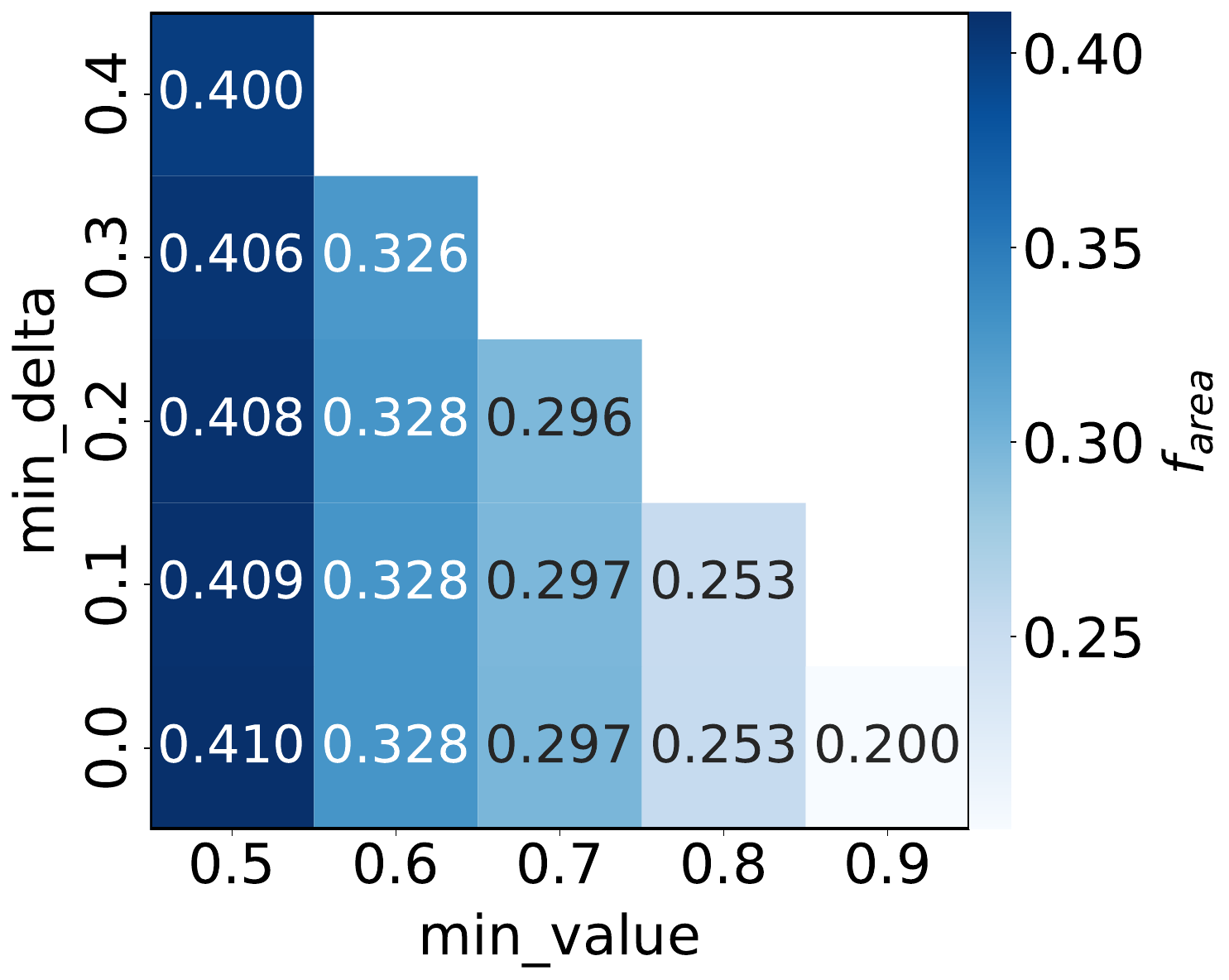}
    \caption{\textit{Left}: Heatmap showing the variation in the number of detected structures (leaves and branches at the base-level) as a function of the \textit{min\_value} and \textit{min\_delta} parameters. The number of structures increases with higher \textit{min\_value} and decreases with increasing \textit{min\_delta}. 
    \textit{Right}: Heatmap illustrating the variation in the fractional area covered by structures (Eq.~\ref{eq:frac_area}). The fractional area covered by detected structures decreases with both increasing \textit{min\_value} and \textit{min\_delta}.}
    \label{fig:heatmaps}
\end{figure}

To illustrate directly how the choice of \textit{min\_delta} affects the identified substructure, Fig.~\ref{fig:min_delta_var} shows the contours of the identified base-level leaves and branches for \textit{min\_delta} = 0, 0.1, 0.2, and 0.25 at fixed \textit{min\_value} = 0.7. As \textit{min\_delta}  increases, the total number of identified structures decreases only mildly (from 307 to 285), but branches are progressively converted into leaves (their number drops from 122 to 13): substructures are absorbed into their parents and the hierarchical information is largely erased, while the outer boundaries of the structures remain essentially unchanged. Approaching the formal limit, we still identify 261 structures for \textit{min\_delta} = 0.29.
Since our analysis relies on the full hierarchy of substructures, and since $R(x,y)$ represents a likelihood rather than a physical intensity, such that a minimum `height' of a substructure in $R$ has no direct physical interpretation; we thus adopt \textit{min\_delta} = 0. The filtering of spurious small-scale fluctuations is then left to \textit{min\_npix}, which imposes a physical cut-off on the smallest structures (see Sect.~\ref{subsec:dendrogram}).

To quantify the coverage by filamentary structures (right panel of Fig.~\ref{fig:heatmaps}) we define two quantities. First is the total area of the map covered by the structures (leaves + branches at the base-level) in the dendrogram, denoted by $A_\textrm{covered}$. The other quantity is the total area of the map covered by the RHT output, denoted by $A_\textrm{map}$. The fractional area is then defined as
\begin{equation}\label{eq:frac_area}
    f_\textrm{area} = \frac{A_\textrm{covered}}{A_\textrm{map}}
.\end{equation}
We find that $f_\textrm{area}$ decreases with both increasing \textit{min\_value} and \textit{min\_delta}, consistent with expectations: as thresholds rise, more diffuse material (low RHT values) is excluded from the analysis.

\begin{figure*}
    \centering
    \subfigure{\includegraphics[width=0.33\linewidth]{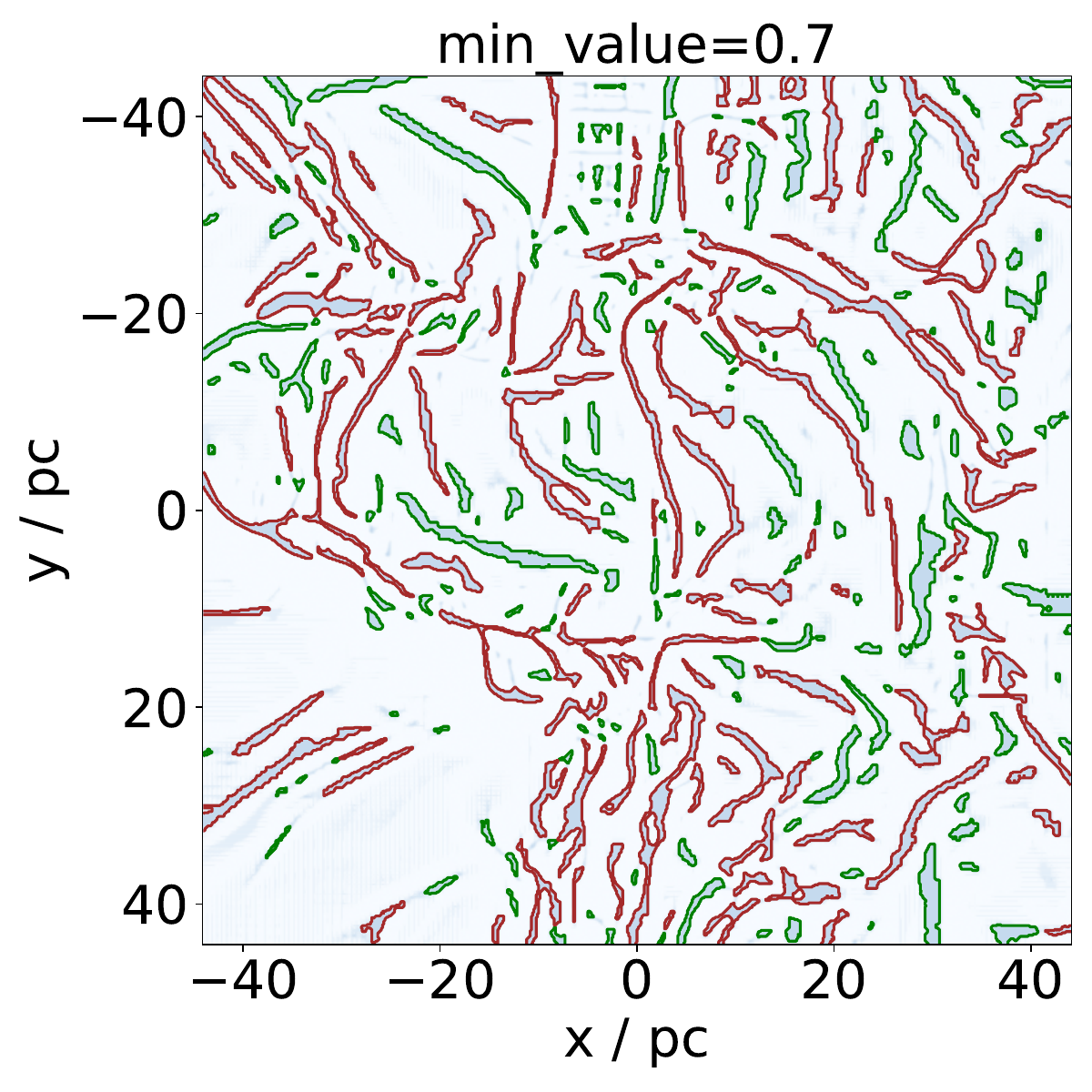}} 
    \subfigure{\includegraphics[width=0.33\linewidth]{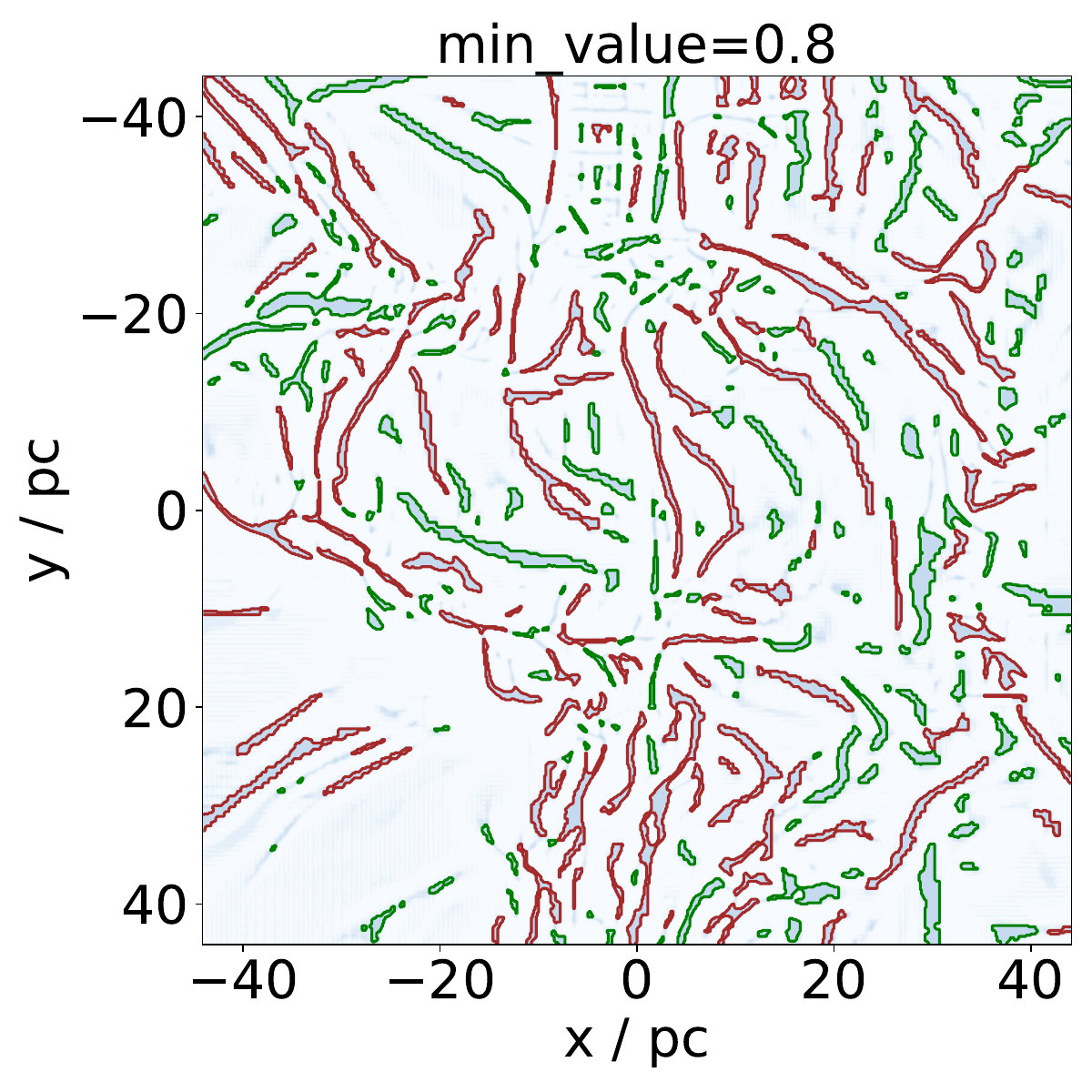}} 
    \subfigure{\includegraphics[width=0.33\linewidth]{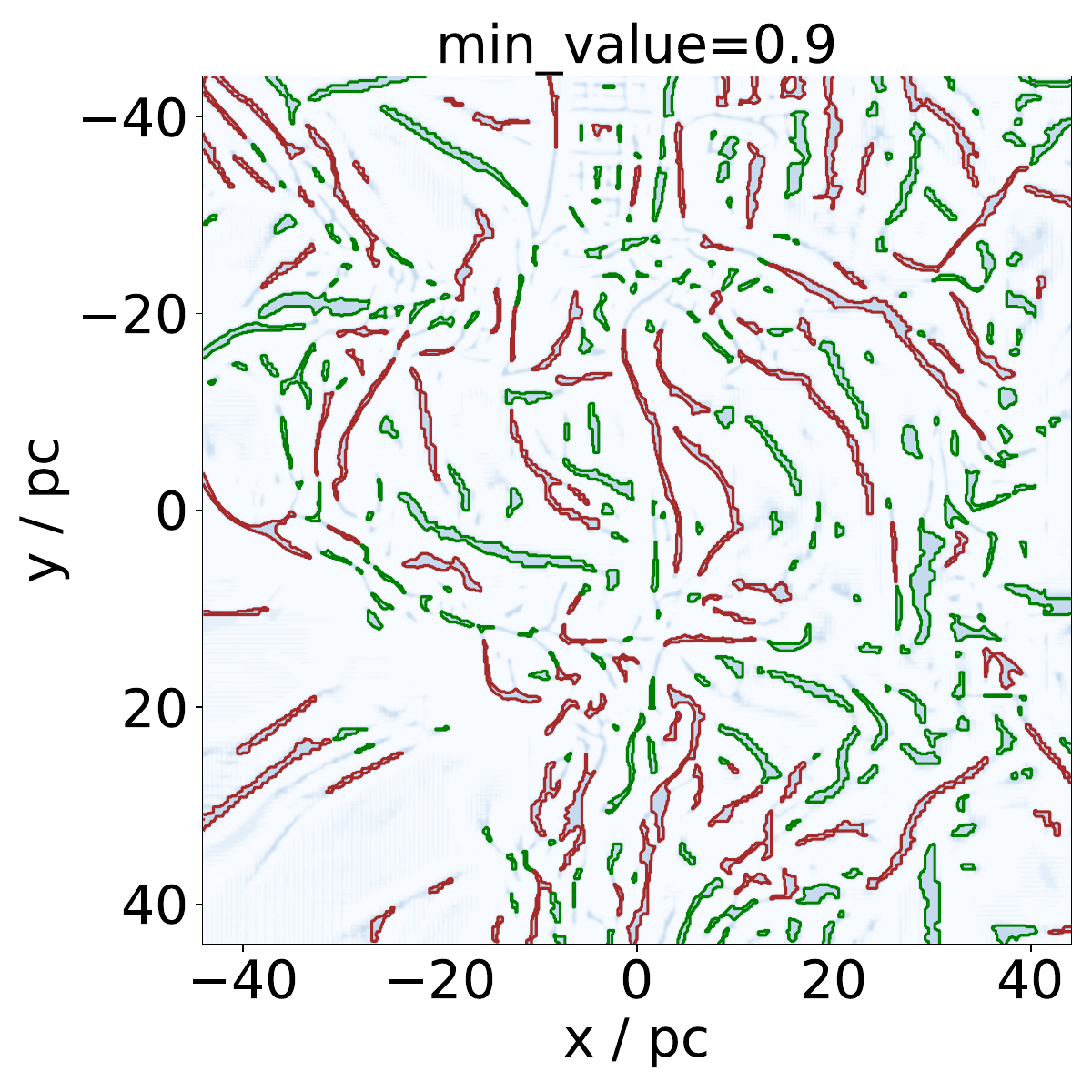}} 
    \caption{Contours of individual structures showing leaves (green) and branches (red) at the base-level identified for different \textit{min\_value} values (increasing from left to right) in the column density map of MC1-HD at $t_\mathrm{evol}$=2.5~Myr projected along the $z$-direction. As \textit{min\_value} increases, the number of detected structures decreases, retaining only the most prominent features. A good balance between capturing real structures and suppressing noise is achieved at \textit{min\_value} = 0.7.}
    \label{fig:min_vals}
\end{figure*}

To preserve as much structural area as possible, we fixed \textit{min\_delta} = 0.0. At the same time, it allowed us to follow the detailed substructure of the identified filaments. 
Furthermore, at very low values of \textit{min\_value} ($\sim$0.5 - 0.6), many independent structures are merged into larger complexes, reducing structural resolution. At the high end (\textit{min\_value} = 0.9), filamentary structure becomes sparse, suggesting the threshold is too restrictive. By comparing structural contours for \textit{min\_value} values in the range [0.7, 0.9] (Fig.~\ref{fig:min_vals}), we find that \textit{min\_value} = 0.7 strikes the best balance, preserving detail while minimizing noise. The combination of parameters as presented in the lower part of Table~\ref{tab:parameters} thus optimally balances the trade-off between detecting genuine filamentary structures and minimizing contamination from artifacts and noise.

\section{Choice of \textit{min\_value} across resolutions}\label{app:choice_of_min_val}

The RHT output intensities vary systematically with the resolution of the underlying column density maps. While the RHT parameters require no modification with changes in map resolution, we find that the threshold parameter \textit{min\_value} of \texttt{AstroDendro} must be adjusted to ensure consistent structure recovery across scales. Without this adjustment, the fractional area (Eq.~\ref{eq:frac_area}) decreases dramatically with resolution -- for instance, by more than 0.1 at a resolution of $\Delta$x = 0.96~pc compared to the \mbox{$\Delta$x = 0.12~pc} case.

To determine an appropriate \textit{min\_value} for lower-resolution maps, we adopted a percentile-matching approach. We began with the highest resolution map ($\Delta$x = 0.12~pc) for which we used \textit{min\_value} = 0.7. We then computed the fraction of pixels in the RHT output that exceeded this threshold, effectively capturing the area covered by significantly linear structures at that resolution. We carried out this procedure for the column density map of MC1-HD at $t_\mathrm{evol} = 2.5$~Myr, projected along the $z$-direction. Since the chosen RHT and \texttt{AstroDendro} parameters (see Appendix~\ref{app:para_opt}) yield consistent results across all column density maps (different clouds, LOS, and evolutionary times), we assume that the resolution-dependent \texttt{AstroDendro} parameter $min\_value$ determined here is likewise applicable to the full set of maps. 

\begin{figure}
    \centering
    \includegraphics[width=1\linewidth]{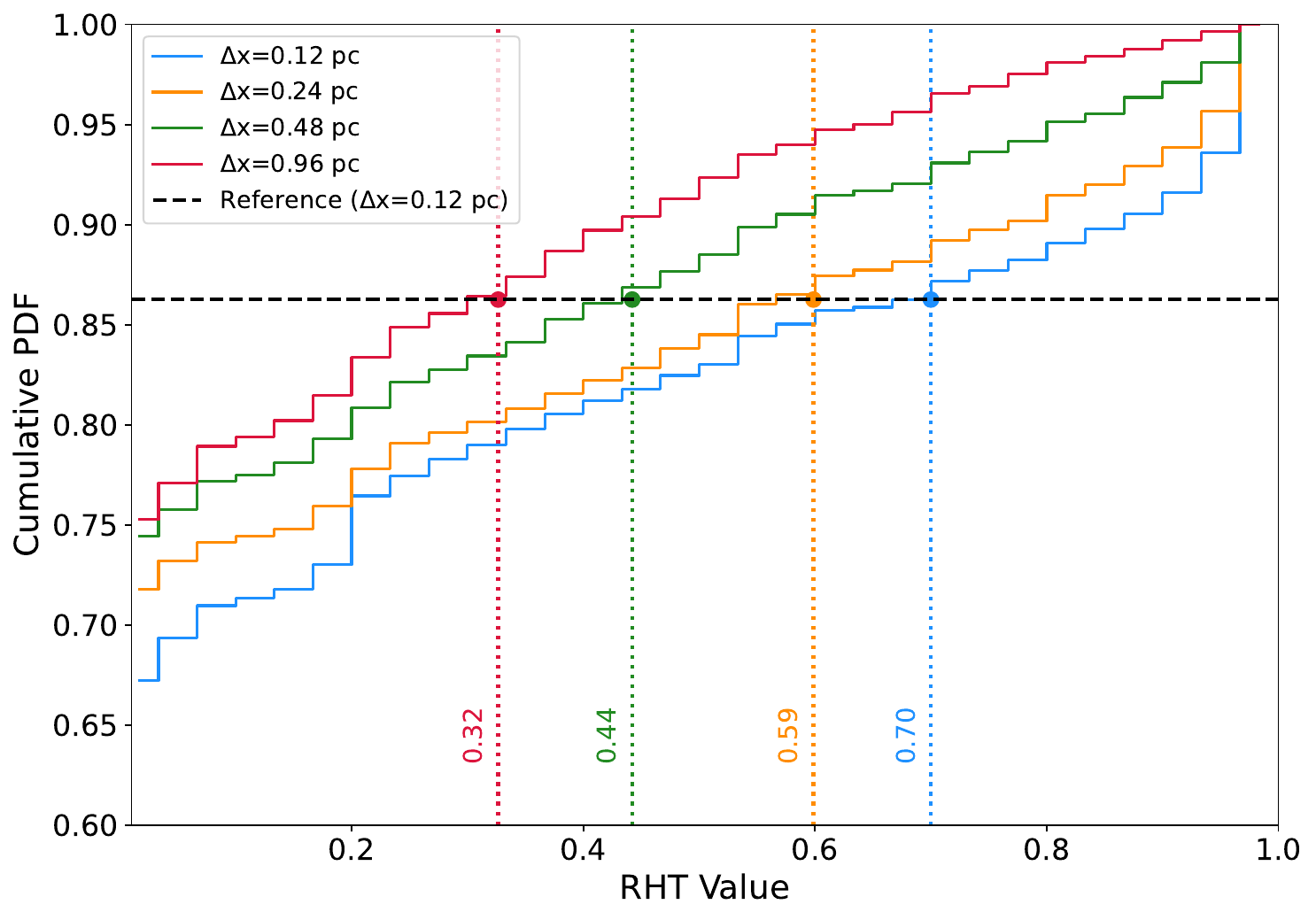}
    \caption{Cumulative PDF of RHT output values for different resolution levels of the column density map of run MC1-HD at $t_\mathrm{evol}$=2.5~Myr projected along the $z$-direction. The horizontal dashed black line represents the reference area fraction corresponding to \textit{min\_value} = 0.7 at the highest resolution (\mbox{$\Delta$x = 0.12 pc}). For each lower resolution, the intersection point between the cumulative PDF and this line is marked with a colored dot. Vertical lines in matching colors pass through these intersection points to indicate the corresponding adapted \textit{min\_value} for the subsequent dendrogram analysis. This approach ensures consistent structural coverage across all resolutions by preserving the same fractional area of filamentary structures.}
    \label{fig:cumulative_histograms}
\end{figure}

Next, we construct cumulative PDFs of the RHT output values at each resolution level of the MC1-HD column density map at $t_\mathrm{evol} = 2.5$~Myr, projected along the $z$-direction. These are shown in Fig.~\ref{fig:cumulative_histograms}. The horizontal dashed black line indicates the reference percentile corresponding to \textit{min\_value} = 0.7 at the highest resolution. For each lower resolution map, we identify the point at which its cumulative PDF intersects this percentile line. The corresponding $x$-axis value at the intersection is chosen as the  \textit{min\_value} for that resolution. The resulting values for \textit{min\_value} are listed in Table~\ref{tab:parameters}.

This method thus ensures that the same fraction of the image area is selected as filamentary structure across resolutions, maintaining good (though not perfect) consistency in structural coverage among the different resolutions and mitigating resolution-induced biases.

\section{Measurement of the filament width}\label{app:width}

\begin{figure}
    \centering
    \includegraphics[width=1\linewidth]{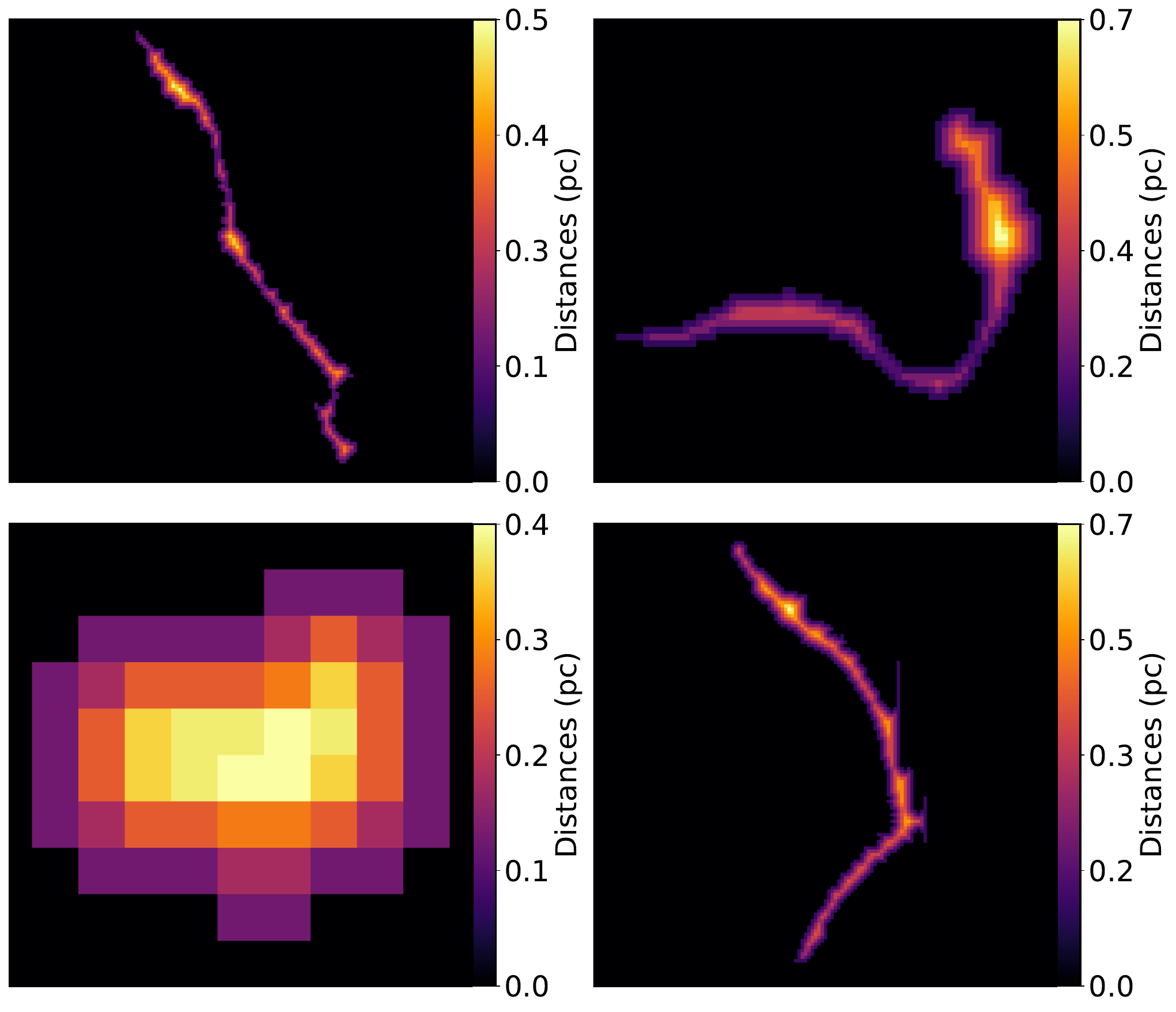}
    \caption{Distance transform maps of some selected filaments. The colormap shows the distance of each pixel to its nearest background pixel. The wider a portion of the filament, the greater is the value at its center.}
    \label{fig:dt_maps}
\end{figure}

To calculate the width of filaments, we depart from the traditional FWHM used in observations and implement an image-processing technique known as the \textit{Distance Transform} (DT), which computes the distance of each pixel in a binary image to the nearest background (`off') pixel, using a chosen metric. We used the Euclidean metric, which measures straight-line distances. The method is implemented using \texttt{distance\_transform\_edt}\footnote{\url{https://docs.scipy.org/doc/scipy/reference/generated/scipy.ndimage.distance_transform_edt.html}} from the \textit{SciPy} Python library, applied directly to the binary filament mask. The resulting distance map assigns each pixel its distance to the nearest background pixel
(see Fig.~\ref{fig:dt_maps}) for a few filamentary structures. Spine pixels (not shown in Fig.~\ref{fig:dt_maps}) that lie along the filament’s center, are then used to estimate the local filament width: taking twice the distance value yields the full width. We then computed the mean of these widths for all pixels along the spine to characterize the filament’s overall width.

\section{Impact of the local background on filament masses and the $L-M$ relation}\label{app:background}

As described in Sect.~\ref{subsec:physical_properties}, the mass of each identified structure is obtained by integrating the raw surface density over its entire area, without subtracting a local background. Here we quantify how this choice affects the masses and the resulting $L-M$ relation. In all tests described below the segmentation itself is left untouched: the structures, and hence their lengths $L$, are exactly those of the main analysis, and only the mass changes. We consider two complementary background estimators, a local one tied to the immediate surroundings of each structure and a global one tied to a fixed spatial scale.

\begin{figure}
    \centering
    \includegraphics[width=1\linewidth]{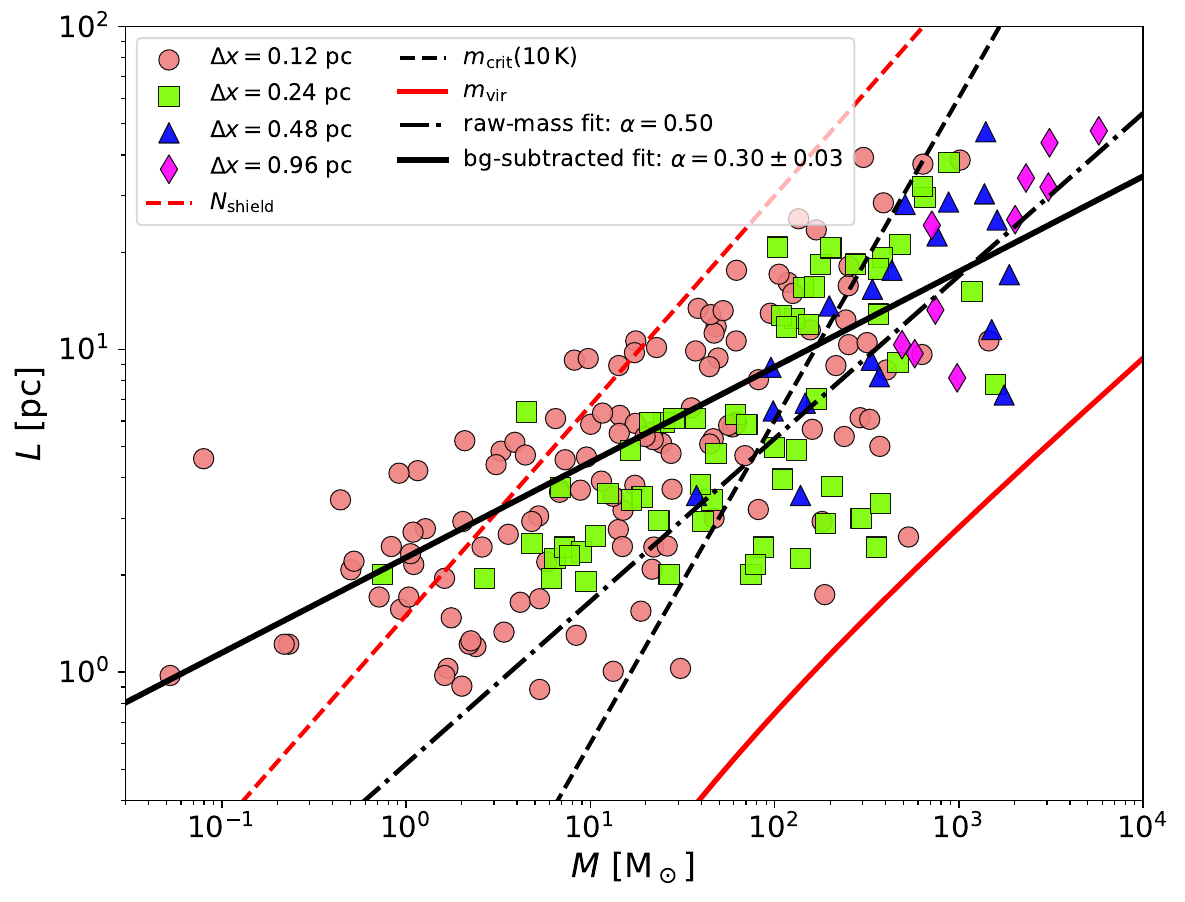}
    \caption{Same as Fig.~\ref{fig:M-L_baselevel_all_res} but for background-subtracted masses (via the local annulus estimator). The solid black line is the power-law fit to the background-subtracted masses ($\alpha = 0.30\pm0.03$). For comparison, the black dash-dotted line shows the fit to the corresponding raw masses ($\alpha = 0.50$) of Fig.~\ref{fig:M-L_baselevel_all_res}. Background subtraction lowers the masses, more so for the smaller structures, which leads to a shallower slope.}
    \label{fig:background}
\end{figure}

Local annulus estimator. Following common observational practice \mbox{\citep[e.g.,][]{schisano_2014}}, we assign to each structure a background pedestal $\Sigma_\mathrm{bg}$, defined as the median surface density in a three-pixel-wide annulus around its mask. The annulus is constructed with \texttt{scipy.ndimage.binary\_dilation} by dilating the structure mask by six pixels and removing from the result the mask dilated by three pixels, so the ring starts three pixels outside the structure; pixels belonging to any other identified structure are excluded. The net mass is then obtained by integrating $\Sigma - \Sigma_\mathrm{bg}$ over the structure's area. We applied this to every map---all clouds, LOS, and evolutionary times at the four resolutions \mbox{$\Delta x$ = 0.12}, 0.24, 0.48, and 0.96~pc, and fit each map exactly as in Sect.~\ref{subsec:filaments_at_multi_res}. The pedestal accounts for a large fraction of the raw integrated mass, with a median of 79\% and a 16th--84th percentile range of 48--95\%. Figure~\ref{fig:background} shows the resulting $L-M$ distribution for the same map as Fig.~\ref{fig:M-L_baselevel_all_res}: background subtraction shifts the structures to significantly lower masses, and it does so more strongly for the smaller structures, which barely exceed their local pedestal. Averaging the fitted power-law slopes (Eq.~\ref{eq:M-L}) over all maps, we obtain $\langle\alpha\rangle = 0.34 \pm 0.04$, compared to $\langle\alpha\rangle = 0.54 \pm 0.07$ for the raw masses of the same structures.

Large-scale top-hat estimator. As a second, deliberately more aggressive definition, we estimate the background by smoothing $\Sigma$ with a circular top-hat kernel and subtracting the smoothed map pixel by pixel, truncating negative values at zero. We used kernel radii of 1.1~pc (the \textit{smr} scale of the RHT unsharp mask, Table~\ref{tab:parameters}), 2.2~pc, and 3.3~pc. Structures comparable in size to, or larger than, the kernel then partly subtract themselves, so this estimator removes even more mass than the annulus: the median background fractions are 87\%, 82\%, and 80\% for the three radii. This test was carried out for MC1-HD at the highest resolution, i.e., for nine maps (three LOS $\times$ three evolutionary times), and yields $\langle\alpha\rangle = 0.33 \pm 0.07$, $0.26 \pm 0.08$, and $0.22 \pm 0.05$ for increasing kernel radius. For the raw masses of these same nine maps we obtain $\langle\alpha\rangle = 0.53 \pm 0.05$, in agreement with the multi-resolution value quoted above, so the two tests can be compared directly: for the smallest kernel, whose radius is closest to the scale probed by the annulus, both estimators give the same slope of $\langle\alpha\rangle \simeq 0.33$, and the slope decreases further as progressively more of the structures' own emission is absorbed into the ``background''.

Uniform background. For completeness, we also tested the simplest conceivable correction: subtracting a spatially uniform background from the surface density map and re-running the full pipeline (RHT, dendrogram segmentation, and characterization) on the subtracted map. Since the unsharp mask underlying the RHT (Sect.~\ref{subsec:RHT}) is invariant under a constant offset, the identification is affected only in the faintest regions, where clipping (unphysical) negative surface densities modifies the map: for the $z$-projection of MC1-HD at $t_\mathrm{evol}$ = 2.5~Myr, subtracting the median diffuse level ($5\times10^{-4}~\mathrm{g\,cm^{-2}}$) leaves 83\% of the structures with a direct counterpart and changes the fitted slope from $\alpha = 0.49 \pm 0.02$ to $0.48 \pm 0.02$. A uniform value, however, cannot represent the actual background.

Taken together, these tests show that the $L-M$ relation remains clearly sub-linear under every background definition we tried, with $\langle\alpha\rangle$ between 0.22 and 0.34. That the slope becomes shallower is expected rather than surprising: background subtraction removes a roughly comparable pedestal from all structures and therefore affects the smaller ones more, stretching the distribution toward low masses at a fixed length. The amount by which $\alpha$ decreases, on the other hand, is set entirely by how much of the surroundings one chooses to call background. As discussed in Sect.~\ref{sec:mass}, this choice cannot be made on the basis of the projected maps alone and in a hierarchically structured medium it is ambiguous also in principle: the background of a substructure partly consists of its parent structures, so setting a background level amounts to setting where one level of the hierarchy is taken to end and the next to begin.
For this reason, and because the observed $L-M$ compilations to which we compare are based on heterogeneous mass estimates, many of which do not subtract a local background \mbox{\citep{hacar_2023}}, we use the raw masses in the main text; $\alpha \simeq 0.5$ is thus the appropriate value for a like-for-like comparison with observations, while the tests above quantify the systematic uncertainty attached to it.

\end{appendix}

\end{document}